\documentclass[%
preprint,showkeys,
 amsmath,amssymb,
 aps,
]{revtex4-1}

\usepackage{graphicx}% Include figure files
\usepackage{dcolumn}% Align table columns on decimal point
\usepackage{bm}% bold math
\begin{document}

\title{Standard and non-standard neutrino-nucleus reactions
  cross sections and event rates to neutrino detection experiments}% Force line breaks 

\author{D.K. Papoulias}
 
  \email{dimpap@cc.uoi.gr}
\author{T.S. Kosmas}%
 \email{hkosmas@uoi.gr}
 \affiliation{Division of Theoretical Physics, University of Ioannina, GR 45100 Ioannina, Greece.}

\begin{abstract}
%%%%%%%%%%%%%%%%%%%%%%%%%%%%%%%%%%%%%%%%%%%%%%%%%%%%%%%%%%%%%%%%%%%
Open neutrino physics issues require precision studies, both theoretical and experimental ones, and towards this aim coherent neutral current neutrino-nucleus scattering events are expected to be observed soon. In this work, we explore $\nu$-nucleus processes from a nuclear theory point of view and obtain results with high confidence  level based on accurate nuclear structure cross sections calculations. Besides cross sections, the present study includes simulated signals expected to be recorded by nuclear detectors, differential event rates as well as total number of events predicted to be measured. Our original cross sections calculations are focused on measurable rates for the Standard Model process, but we also perform calculations for various channels of the non-standard neutrino-nucleus reactions and come out with promising results within the current upper limits of the corresponding exotic parameters.  We concentrate on the possibility of detecting (i) supernova neutrinos by using massive detectors like those of the GERDA and SuperCDMS dark matter experiments and (ii) laboratory neutrinos produced near the spallation neutron source facilities (at Oak Ridge National Lab) by the COHERENT experiment. Our nuclear calculations take advantage of the relevant experimental sensitivity and employ the severe bounds extracted for the exotic parameters entering the Lagrangians of various particle physics models and specifically those resulting from the charged lepton flavour violating $\mu^{-} \rightarrow e^{-}$ experiments (Mu2e and COMET experiments).
%%%%%%%%%%%%%%%%%%%%%%%%%%%%%%%%%%%%%%%%%%%%%%%%%%%%%%%%%%%%%%%%%%%

\end{abstract}


\maketitle

\section{Introduction}
\label{Intro}

Coherent scattering of neutrinos on complex nuclei was proposed long ago \cite{Freedman,Drukier-Stodolsky} as a prominent probe to study  neutral-current (NC) $\nu$-nucleus processes, but up to now no events have been experimentally measured. Neutrino detection, constitutes an excellent probe to search for a plethora of conventional neutrino physics applications and new-physics open issues \cite{Langanke,Ejiri,Kosm-Oset}. In principle, low-energy astrophysical and laboratory neutrino searches provide crucial information towards understanding the underling physics of the fundamental electroweak interactions within and beyond the SM \cite{Schechter-Valle}. Well-known neutrino sources include (i)  supernova neutrinos (with energies up to 60-100 MeV) and (ii) laboratory neutrinos (with energies up to 52.8 MeV) emerging from stopped-pion and muon decays at muon factories (Fermilab, PSI, JPARC, etc.) and at the spallation neutron source (SNS) at Oak Ridge National Lab \cite{SNS}. Recently, it became feasible \cite{Scholberg} to detect neutrinos by exploiting the neutral current interactions and measuring the nuclear recoil signals through the use of very low threshold-energy detectors \cite{pion-DAR-nu,Louis}. To this purpose, great experimental effort has been put and new experiments have been proposed to be performed at facilities with stopped-pion neutrino beams, based on promising nuclear detectors like those of the COHERENT experiment \cite{coherent1,coherent2} and others \cite{CLEAR} at the SNS, or alternative setups at the Booster Neutrino Beam (BNB) at Fermilab \cite{BNB,Brice}. The  nuclear $\nu$-detectors adopted by the relevant experiments include liquid noble gases, such as $^{20}$Ne, $^{40}$Ar, $^{132}$Xe as well as, $^{76}$Ge and CsI[Na] detection materials \cite{Collar}.

On the theoretical side, the $\nu$-signals of low-energy neutrinos, expected to be recorded in sensitive nuclear detectors \cite{Hirata-Bionta,Keil}, could be simulated through nuclear calculations of $\nu$-nucleus scattering cross sections. Such results may provide  useful information relevant for the evolution of distant stars, the core collapse supernovae, explosive nucleosynthesis and other phenomena \cite{Haxton,Giannaka}. In fact, coherent neutral current $\nu$-nucleus scattering events are expected to be observed by using the high intensity stopped-pion neutrino beams \cite{Avignone-Efremenko,Efremenko-2009} and nuclear targets for which recoil energies are of the order of a few to tens of keV, and therefore appropriate for detection of WIMPs \cite{Horowitz,Anderson}, candidates of cold dark-matter \cite{Kosmas-dark_matter,Kosmas-cdm}. Such detectors are e.g. the SuperCDMS \cite{SuperCDMS}, GERDA \cite{GERDA} and other multi-purpose detectors \cite{CLEAN,WARP,XENON-100}. For low-energies, the dominant vector components of NC interactions lead to a coherent contribution of all nucleons (actually all neutrons) in the target nucleus \cite{Giom-Vergados,Monroe_Fischer,Biassoni}. 

It is worth mentioning that, after the discovery \cite{SK,SNO,KamLAND,Tortola,Valle} of neutrino oscillations in propagation, the challenge of neutral and charged lepton flavour violation (LFV) is further investigated by  extremely sensitive experiments \cite{COMET,Mu2e,Bernstein-Cooper,PRIME,Kuno-PRIME} searching for physics beyond the current Standard Model (SM) \cite{Kuno}. To this end, neutrino-nucleus coherent scattering experiments may probe new physics beyond the SM involved in exotic neutrino-nucleus interactions \cite{Scholberg,Davidson,Barranco,PLB}, an undoubtable signature of non-standard physics. Therefore, new data and insights will be provided to the physics of flavour changing neutral current (FCNC) processes, in the leptonic sector, in non-standard neutrino oscillation effects \cite{Friedland-solar,Friedland-atm,Friedland-atm2}, in neutrino transition magnetic moments \cite{Healey}, in sterile neutrino search \cite{sterile} and others \cite{Amanik-McLaughin}. Furthermore, such experimental sensitivity may also inspire advantageous probes to shed light on various open issues in nuclear astrophysics \cite{Amanik_2005,Tomas-Valle-10}. 

In recent works \cite{PLB}, neutral-current (NC) non-standard interactions (NSI) involving (anti)neutrino scattering processes on leptons, nucleons and nuclei have been investigated. The reactions of this type that take place in nuclei are  represented by 
\begin{equation}
 \nu_{\alpha}(\tilde{\nu}_{\alpha}) + (A,Z) \rightarrow \nu_{\beta}(\tilde{\nu}_{\beta})  + (A,Z) \, ,
\label{neutrin-NSI}
\end{equation} 
($\alpha, \beta = e,\mu,\tau$ with $\alpha \neq \beta$). It has been suggested \cite{Papoulias 410} that, theoretically the latter processes, can be studied with the same nuclear methods as the exotic cLFV process of $\mu^-\to e^-$ conversion in nuclei \cite{Chiang-Kosm,Kosm_PhysRep,Kosmas_A683}. The corresponding Lagrangians may be derived within the context of various extensions of the SM \cite{Schechter-Valle,Kosm-Leonta-Vergad}, like the four fermion contact interaction, seesaw model \cite{Kos_Dep_Wal,Forero}, left-right symmetric models \cite{Dep-Valle}, gluonic operator model \cite{Zhuridov}, etc.
 
It is well-known that neutrino NSI may have rather significant impact in many areas of modern physics research and thus, motivate a great number of similar studies \cite{Amanik_2007}. Especially in astrophysical applications, constraints coming out of some supernova explosion scenarios \cite{Barranco-Walle,Mir-Tort-Walle,Bar-Mir-Rashba}, may be affected and eventually lead to the necessity of further investigation of NSI in both LFV and cLFV processes that may occur in solar and supernova environment \cite{Kosm_B215,Kosm-A570,Kos_Kov_Schm}. Such open issues motivated our present work too. 
 
One of our main purposes in this paper, which is an extension of our previous study \cite{PLB}, is to comprehensively study the above issues by performing nuclear structure calculations for a set of experimentally interesting nuclei. We estimate reliably the nuclear matrix elements describing both interaction channels, the exotic and the Standard Model ones, but we mainly focus on the SM component of the neutrino-nucleus processes, i.e. we consider $\alpha=\beta$ in the reactions of Eq. (\ref{neutrin-NSI}). Exotic neutrino-nucleus events are also computed. By exploiting our accurate original cross sections,  we obtain simulated $\nu$-signals and flux averaged cross sections  which are experimentally interesting quantities for both Supernova and SNS neutrinos. The total number of events expected to be recorded over the energy-threshold for the studied nuclear targets are also presented for both cases. 

We stress that, we have devoted special effort to obtain results of high accuracy by constructing the nuclear ground state within the context of the quasi-particle random phase approximation (QRPA), i.e. by solving iteratively the BCS equations for realistic pairing interactions (the Bonn C-D potential) \cite{Chassioti_2009,vtsak-tsk-1}, and achieving high reproducibility of the available experimental data \cite{deVries}. In addition, we made comparisons with the results of other methods evaluating the nuclear form factors that enter the coherent rate \cite{Don-Wal,Alberico} as the one which employs fractional occupation probabilities (FOP) of the states (on the basis of analytic expressions) \cite{Kosm_A536},  and other well-known methods \cite{Engel}. 
%%%%%%%%%%%%%%%%%%%%%%%%%%%%%%%%%%%%%%%%%%%%%%%%%%%%%%%%%%%%%%%%%%%%%%%%%%%%%

\section{Description of the formalism}
\label{chapt2}

In this section we present briefly the necessary formalism for describing all channels of the NSI processes of the reactions (\ref{neutrin-NSI}), derived by starting from the corresponding nuclear-level Feynman diagrams.
%
%%%%%%%%%%%%%%%%%%%%%%%%%%%%%%%%%%%%%%%%%%%%%%%%%%%%%%%%%%%%%%%%%%% 
\begin{figure}[ht]
\begin{center}
\includegraphics[width=0.65\textwidth]{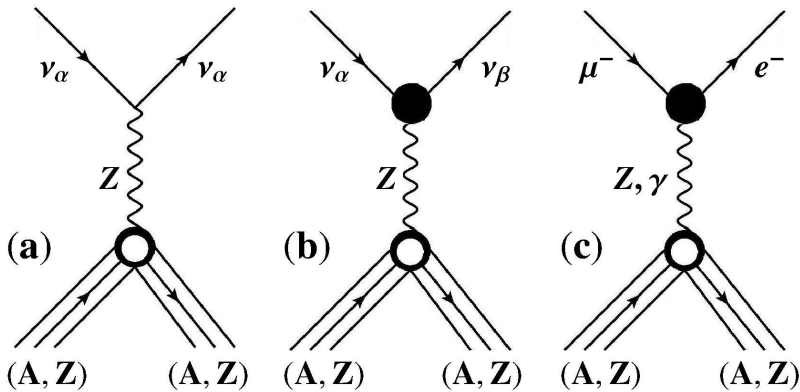}
\end{center}
\caption{ Nuclear level Feynman diagrams for: (\textit{a}) SM Z-exchange neutral current $\nu$-nucleus 
reactions, (\textit{b}) non-standard Z-exchange $\nu$-nucleus reactions, and (\textit{c}) Z-exchange 
and photon-exchange $\mu^{-}\rightarrow e^{-}$ in the presence of a nucleus (muon-to-electron conversion). 
The non-standard (cLFV or LFV) physics enters in the complicated vertex denoted by the bullet $\bullet$ \cite{PLB}.} 
\label{fig.1}
\end{figure}
%%%%%%%%%%%%%%%%%%%%%%%%%%%%%%%%%%%%%%%%%%%%%%%%%%%%%%%%%%%%%%%%%%% 
%
In Fig. \ref{fig.1} the exchange of a $Z$-boson between a lepton and a nucleon is represented, for the SM $\nu$-nucleus scattering, Fig. \ref{fig.1}(a), and for the exotic $\nu$-nucleus scattering, Fig. \ref{fig.1}(b). As already mentioned in the Introduction, the non-standard $\nu$-nucleus processes \cite{PLB}  and the exotic cLFV $\mu^-\to e^-$ conversion 
in nuclei \cite{Kuno,Chiang-Kosm,Kosm_B215,Kos_Kov_Schm}, can be predicted within the context of the same new-physics models \cite{Papoulias 410,Kos_Dep_Wal}. For this reason, in Fig. \ref{fig.1}(c) we also show the exchange of a $Z$-boson or a virtual $\gamma$-photon leading to the nuclear $\mu^-\to e^-$ conversion, 
\cite{Kosm_PhysRep,Kos_Dep_Wal}. Thus, the leptonic vertex in the cases of Fig. \ref{fig.1}(b),(c) is a 
complicated one. A general effective Lagrangian that involves SM interactions ($\mathcal{L}_{\mathrm{SM}}$) 
and NSI ($\mathcal{L}_{\mathrm{NSI}}$) with a non-standard flavour preserving (FP) term, a non-universal (NU) term and a flavour changing (FC) term reads
\begin{equation}
\mathcal{L}_{\mathrm{tot}} = \mathcal{L}_{\mathrm{SM}} + \mathcal{L}_{\mathrm{NSI}} = 
     \mathcal{L}_{\mathrm{SM}} + \mathcal{L}_{\mathrm{NU}} + \mathcal{L}_{\mathrm{FC}} \, . 
\label{tot_Lagr}
\end{equation}
Each of the components $\mathcal{L}_{\mathrm{SM}}$ and $\mathcal{L}_{\mathrm{NSI}}$, the individual terms $\mathcal{L}_{NU}$ and $\mathcal{L}_{FC}$ as well as the nuclear matrix elements that arise from each part, are discussed below. 

\subsection{Coherent cross sections of non-standard $\nu$-nucleus reactions}

The quark-level Lagrangian for neutral current non-standard neutrino interactions $\mathcal{L}_{\mathrm{NSI}}$, at the four fermion approximation, (energies $\ll M_{Z}$) is parametrized  as \cite{Barranco,Barranco-Walle,Scholberg} 
\begin{equation}
\mathcal{L}_{\mathrm{NSI}} = - 2\sqrt{2} G_F \sum_{\begin{subarray}{c} f= \, u,d\\ \alpha,\beta = \, e,\mu,\tau\end{subarray}} 
\epsilon_{\alpha \beta}^{f P}\left[\bar{\nu}_{\alpha}\gamma_\rho L \nu_\beta\right]\left[\bar{f}\gamma^\rho P f\right]\, ,
\label{NSI_Lagr}
\end{equation}
where $f$ denotes a first generation SM quark, $\nu_{\alpha}$ are three light neutrinos  with Majorana masses and $P=\lbrace L, R\rbrace$ are the chiral projectors. In the latter Lagrangian (\ref{NSI_Lagr}), two classes of non-standard terms are considered (i) flavour preserving non-SM terms that are proportional to $\epsilon_{\alpha \alpha}^{f P}$ (known as non-universal, NU interactions) and (ii) flavour-changing (FC) terms proportional 
to $\epsilon_{\alpha \beta}^{f P}$, $\alpha\neq\beta$.
 These couplings are defined with respect to the strength 
of the Fermi coupling constant $G_F$ \cite{Barranco,Barranco-Walle}. In the present work we examine spin-zero nuclei thus, the polar-vector couplings defined as $\epsilon_{\alpha\beta}^{f V}=\epsilon_{\alpha\beta}^{f L} + \epsilon_{\alpha\beta}^{f R}$ are mainly of interest. For the axial-vector couplings it holds $\epsilon_{\alpha\beta}^{f A}=\epsilon_{\alpha \beta}^{f L} - \epsilon_{\alpha\beta}^{f R}$. 

Following  Ref. \cite{Kos_Kov_Schm}, the nuclear physics aspects of the neutrino-matter NSI can be explored by transforming the quark-level Lagrangian (\ref{NSI_Lagr}), eventually to the nuclear level where the hadronic current is written in terms of NC nucleon form factors that are functions of the four momentum transfer. Generally, for  inelastic $\nu$-nucleus scattering, the magnitude of the three momentum transfer, $q = \vert{\vec{q}}\vert$, is a function of the scattering angle of the outgoing neutrino $\theta$ 
(in laboratory frame), the initial, $E_{i}$, and final, $E_{f}$, nuclear energies, as well as the excitation energy of the target nucleus, $\omega$, and takes the form $q^2=\omega^2+ 2 E_{i} E_{f} \left(1 - \cos \theta \right)$ \cite{Chassioti_2009,Don-Wal}.
Our analysis in the present paper, concentrates on the dominant coherent (elastic) channel where only $gs \rightarrow gs$ transitions occur ($\omega=0$, $E_i = E_f$) and the momentum transfer in terms of the incoming neutrino energy, $E_\nu$, becomes $q^2 = 2 E_\nu^2 (1-\cos \theta)$ or equivalently $q = 2 E_\nu \sin (\theta/2)$.  

The NSI coherent differential cross section of neutrinos scattering off a spin-zero nucleus, with respect to the scattering angle $\theta$  reads \cite{PLB}
\begin{equation}
\frac{d \sigma_{\mathrm{NSI},\nu_{\alpha}}}{d \cos \theta} = \frac{G_{F}^{2}}{2 \pi} E_{\nu}^{2} \left(1 + 
\cos \theta \right)\left\vert 
\langle gs \vert \vert G_{V,\nu_{\alpha}}^{\mathrm{NSI}}(q) \vert \vert gs \rangle \right \vert ^{2},
\label{NSI_dcostheta}
\end{equation}
$\alpha = e,\mu,\tau$, denotes the flavour of incident neutrinos and $\vert gs \rangle$ represents 
the nuclear ground state (for even-even nuclei assumed here, $\vert gs \rangle=\vert J^\pi \rangle\equiv\vert 0^+ \rangle$). The nuclear matrix element, 
that enters the cross section of Eq. (\ref{NSI_dcostheta}), is written as \cite{PLB}
\begin{equation}
\begin{aligned}
& \left\vert {\cal M}^{\mathrm{NSI}}_{V,\nu_{\alpha}} \right \vert ^{2} \equiv
\left\vert \langle gs \vert \vert G_{V,\nu_{\alpha}}^{\mathrm{NSI}}(q)  \vert \vert gs \rangle \right \vert ^{2}  = \\  & \qquad \quad
 \left[ \left( 2 \epsilon_{\alpha \alpha}^{uV} + \epsilon_{\alpha \alpha}^{dV} \right) Z F_Z (q^2) +  
\left( \epsilon_{\alpha\alpha}^{uV} + 2\epsilon_{\alpha\alpha}^{dV} \right) N F_N (q^2) \right]^2 \\
& \, \, \, \, + \sum_{\beta \neq \alpha} \left[\left( 2 \epsilon_{\alpha \beta}^{uV}+ \epsilon_{\alpha \beta}^{dV} 
\right) Z F_Z (q^2)+ \left(\epsilon_{\alpha \beta}^{uV}+ 2 \epsilon_{\alpha\beta}^{dV} \right)N F_N (q^2)\right]^2,
\end{aligned}
\label{GV}
\end{equation}
($\beta = e,\mu,\tau$) where $F_{Z(N)}$ denote the nuclear (electromagnetic) form factors for protons 
(neutrons). We stress that, in the adopted NSI model, the coherent NC $\nu$-nucleus 
cross section is not flavour blind as in the SM case. Obviously, by incorporating the nuclear structure details, in Eqs. (\ref{NSI_dcostheta}) and (\ref{GV}), the cross sections become more realistic and accurate \cite{Scholberg}. The structure of the Lagrangian (\ref{tot_Lagr}), implies that in the r.h.s. of Eq. (\ref{GV}) the first term  
is the NU matrix element, ${\cal M}^{\mathrm{NU}}_{V,\nu_{\alpha}}$, and the summation is the FC 
matrix element, ${\cal M}^{\mathrm{NU}}_{V,\nu_{\alpha}}$, hence we write
\begin{equation}
\left\vert {\cal M}^{\mathrm{NSI}}_{V,\nu_{\alpha}} \right \vert ^{2} = 
\left\vert {\cal M}^{\mathrm{NU}}_{V,\nu_{\alpha}} \right \vert ^{2} 
+ \left\vert {\cal M}^{\mathrm{FC}}_{V,\nu_{\alpha}} \right \vert ^{2} \, . \label{NU-FC-terms}
\end{equation}

From experimental physics perspectives, it is rather crucial to express the differential cross section with respect to the recoil energy of the nuclear target, $T_N$. In recent years it became feasible for terrestrial neutrino detectors to detect neutrino events by measuring nuclear recoil \cite{Brice,Collar}. Therefore, it is  important 
to compute also the differential cross sections $d\sigma/dT_N$. In the  coherent process, the nucleus 
recoils (intrinsically it remains unchanged) with energy which, in the approximation $T_{N} \ll E_{\nu}$ 
takes the maximum value $T_N^{\text{max}}=2 E_\nu^2/(M+ 2 E_\nu)$, with $M$ denoting the nuclear mass 
\cite{Giom-Vergados,Monroe_Fischer}. Then, to a good approximation, the square of the three momentum transfer, 
is equal to $q^2 = 2 M T_N$, and the coherent NSI differential cross section with respect to $T_{N}$ can be cast in the form
\begin{equation}
\frac{d\sigma_{\mathrm{NSI},\nu_{\alpha}}}{dT_N} = \frac{G_F^2 \,M}{\pi} \left(1- 
\frac{M\,T_N}{2 E_\nu^2}\right)\left\vert\langle gs\vert\vert 
G_{V,\nu_{\alpha}}^{\mathrm{NSI}} (q) \vert\vert gs \rangle \right \vert ^{2}\, .
\label{NSI_dT}
\end{equation}
 We note that, compared to previous studies \cite{Amanik_2005,Amanik_2007}, 
we have also taken into consideration the interaction $\nu- u$ quark  [see Eq. (\ref{GV})], in addition to the momentum dependence of the nuclear form factors \cite{PLB}. Both Eqs. (\ref{NSI_dcostheta}) and (\ref{NSI_dT}) are useful for studying the nuclear physics of NSI of 
neutrinos with matter.
 
Furthermore, by performing numerical integrations to Eq.     (\ref{NSI_dcostheta}) over the scattering 
angle $\theta$ or to Eq. (\ref{NSI_dT}) over the recoil energy $T_N$, one can obtain integrated (total) 
coherent NSI cross sections, $\sigma_{\mathrm{NSI},\nu_\alpha}$. Following Eq. (\ref{NU-FC-terms}), the 
individual cross sections $\sigma_{\mathrm{NU,\nu_{\alpha}}}$ and $\sigma_{\mathrm{FC,\nu_{\alpha}}}$ 
may be evaluated accordingly. 
% $\sigma_{\mathrm{NSI,\nu_{\alpha}}} = \sigma_{\mathrm{NU,\nu_{\alpha}}} + \sigma_{\mathrm{FC,\nu_{\alpha}}}$. 
% Results for the $\sigma_{\mathrm{tot}}$ and partial cross sections $\sigma_{\mathrm{NU}}$ 
% and $\sigma_{\mathrm{FC}}$ are  presented and discussed in Section 4.

%%%%%%%%%%%%%%%%%%%%%%%%%%%%%%%%%%%%%%%%%%%%%%%%%%%%%%%%%%%%%%%%%%%%%%%
\subsection{Standard Model coherent $\nu$-nucleus cross sections}

The effective (quark-level) SM $\nu$-nucleus interaction Lagrangian, $\mathcal{L}_{\mathrm{SM}}$ at low and intermediate neutrino energies, is written as
\begin{equation}
\mathcal{L}_{\mathrm{SM}} = - 2 \sqrt{2} G_{F} \sum_{ \begin{subarray}{c} f= \, u,d \\ \alpha= e, \mu,\tau 
\end{subarray}} g_P^f   
\left[ \bar{\nu}_{\alpha} \gamma_{\rho} L \nu_{\alpha} \right] \left[ \bar{f} \gamma^{\rho} P f \right] \, ,
\label{SM_Lagr}
\end{equation}
where $g_L^u = \frac{1}{2}-\frac{2}{3} \sin^2 \theta_W$ and $g_R^u = -\frac{2}{3} \sin^2 \theta_W$ are the 
left- and right-handed couplings of the $u$-quark to the $Z$-boson and $g_L^d = -\frac{1}{2}+\frac{1}{3} \sin^2 \theta_W$ 
and $g_R^d = \frac{1}{3}\sin^2 \theta_W$ are the corresponding couplings of the $d$-quark ($\theta_W$ is the Weinberg 
mixing angle) \cite{{Don-Wal}}.

For coherent $\nu$-nucleus scattering, the SM angle-differential cross section reads
\begin{equation}
\frac{d \sigma_{\mathrm{SM},\nu_{\alpha}}}{d \cos \theta} = \frac{G_{F}^{2}}{2 \pi} E_{\nu}^{2} \left(1 + 
\cos \theta \right)\left\vert\langle gs \vert\vert \hat{\mathcal{M}}_0(q) \vert\vert gs \rangle\right \vert^2\, .
\label{SM_dcostheta}
\end{equation}
The operator $\hat{\mathcal{M}}_0$ in the nuclear matrix element of the latter equation is the Coulomb operator which is equal to the product of the zero-order spherical Bessel function times the zero-order 
spherical harmonic \cite{Chassioti_2009,Don-Wal}. This matrix element can be cast in the form \cite{Kosm-A570}
\begin{equation}
\left\vert {\cal M}^{\mathrm{SM}}_{V,\nu_{\alpha}} \right \vert ^{2} \, \equiv \,
\left\vert\langle gs \vert\vert \hat{\mathcal{M}}_0(q) \vert\vert gs \rangle\right \vert^2 = 
\left[g^p_V Z F_Z (q^2) + g^n_V N F_N (q^2) \right]^2 \, ,
\label{SM-ME}
\end{equation} 
where, the polar-vector couplings of protons $g^p_V$ and neutrons $g^n_V$ with the $Z$ boson (see 
Fig. \ref{fig.1}(a)), are written as 
$g^p_V = 2(g_L^u + g_R^u) + (g_L^d + g_R^d) = \frac{1}{2}-2 \sin^2 \theta_W$ and $g^n_V = (g_L^u + g_R^u)
+2(g_L^d + g_R^d)=-\frac{1}{2}$, respectively. As can be easily seen,  the vector contribution of all protons is very small ($g^p_V \sim 0.04$), hence the coherence in Eq. (\ref{SM-ME}) essentially refers to all neutrons only of the studied nucleus. After some straightforward elaboration the differential cross section with respect to the nuclear recoil energy, $T_N$, takes the form
\begin{equation}
\frac{d\sigma_{\mathrm{SM},\nu_{\alpha}}}{dT_N} = \frac{G_F^2 \,M}{\pi} \left(1- 
\frac{M\,T_N}{2 E_\nu^2}\right)\left\vert\langle gs \vert\vert \hat{\mathcal{M}}_0(q) \vert\vert gs \rangle\right \vert^2\, .
\label{SM_dT}
\end{equation}

The Lagrangian $\mathcal{L}_{\mathrm{tot}}$ of Eq. (\ref{tot_Lagr}),   contains the flavour preserving (FP) part, equal to
$\mathcal{L}_{\mathrm{FP}} \equiv \mathcal{L}_{\mathrm{NU}}+\mathcal{L}_{\mathrm{SM}} $, which can be evaluated through the Coulomb matrix element 
\begin{equation}
\left\vert {\cal M}^{\mathrm{FP}}_{V,\nu_\alpha}\right\vert^2 =\left\vert{\cal M}^{\mathrm{SM}}_{V,\nu_\alpha}  
+  {\cal M}^{\mathrm{NU}}_{V,\nu_{\alpha}} \right \vert ^{2} \, .  
\label{FP}
\end{equation}
Subsequently, the total coherent cross section may  be computed on the basis of the matrix element
\begin{equation}
\left\vert {\cal M}^{\mathrm{tot}}_{V,\nu_\alpha}\right\vert^2 = \left\vert{\cal M}^{\mathrm{FP}}_{V}\right\vert^2 
+  \left\vert{\cal M}^{\mathrm{FC}}_{V,\nu_\alpha}\right\vert^2 \, .  
\label{tot}
\end{equation}

In a previous work \cite{PLB} we evaluated original differential cross sections $d\sigma_{\lambda,\nu_\alpha}/d\cos\theta$ 
and $d\sigma_{\lambda,\nu_{\alpha}}/dT_N$, as well as individual angle-integrated cross sections of the 
form $\sigma_{\lambda,\nu_\alpha} (E_\nu)$, with $\alpha = e,\mu,\tau$, and 
$\lambda= \mathrm{SM, NU, FP, FC}$ (FC stands for the six flavour changing processes
$\nu_e\leftrightarrow\nu_\mu, \, \nu_e\leftrightarrow\nu_\tau, \, \nu_\mu\leftrightarrow\nu_\tau$).  

In this work, we perform Standard model cross sections calculations (for convenience from now on we drop the index $\lambda=\mathrm{SM}$ and always consider $\nu_{\alpha}=\nu_{\beta}$) for a set of nuclei throughout the periodic table up to $^{208}$Pb. We adopt various nuclear models (see section \ref{nucl-methods}) to compute the nuclear form factors. Then, for a great part of the cross section results (except differential cross sections) we evaluate folded cross sections, and event rates.

%%%%%%%%%%%%%%%%%%%%%%%%%%%%%%%%%%%%%%%%%%%%%%%%%%%%%%%%%%%%%%%%%%%%

\section{Evaluation of the nuclear form factors}
\label{nucl-methods}
%%%%%%%%%%%%%%%%%%%%%%%%%%%%%%%%%%%%%%%%%%%%%%%%%%%%%%%%%%%%%%%%%%%%
\subsection{Nuclear Structure calculations}

At first, we study the nuclear structure details of the matrix elements entering 
Eq. (\ref{SM-ME}), such results reflect the dependence of the coherent cross section on the incident-neutrino energy $E_{\nu}$ and the scattering angle $\theta$ (or the recoil energy $T_{N}$). We mention that for the even-even nuclei  this study involves realistic 
QRPA calculations for the differential cross sections $d\sigma_{\nu_{\alpha}}/d\cos\theta$ and $d\sigma_{\nu_{\alpha}}/dT_N$, performed after
constructing the nuclear ground state $\vert gs\rangle$ by solving iteratively the Bardeen Cooper Schriffer (BCS) equations. The solution of these equations provides the probability amplitudes $\upsilon^j_{N_{n}}$ and $\mathrm{v}^j_{N_{n}}$ of the $j$-th single nucleon level to be occupied or unoccupied, respectively. Moreover, the latter equations provide the single quasi-particle energies, based on the single particle energies of the nuclear field (a Coulomb corrected Woods-Saxon potential in our case) as well as the pairing part of the residual two-body interaction (Bonn C-D potential in our case).  Then, the nuclear form factors for protons (neutrons) are obtained as \cite{Kosm-A570}
\begin{equation}
F_{N_n}(q^2) = \frac{1}{N_n}\sum_j [j]\, \langle j\vert j_0(qr)\vert j\rangle\left(\upsilon^j_{N_n}\right)^2 \, ,
\end{equation}
with $[j]=\sqrt{2 j +1}$, $N_{n}=Z \,\,(\mathrm{or}\,\, N)$. For each nuclear system studied, the chosen active model space, the harmonic oscillator (h.o.) parameter $b$ and the values of the two parameters $g^{p\,(n)}_{\mathrm{pair}}$  for proton (neutron) pairs that renormalise the monopole (pairing) residual interaction (obtained from the Bonn C-D two-body potential describing the strong two-nucleon forces), are presented in Table \ref{table1}. The adjustment of $g^{p\,(n)}_{\mathrm{pair}}$ is achieved through the reproducibility of the pairing gaps $\Delta_{p\,(n)}$ (see e.g. \cite{Giannaka}).

%%%%%%%%%%%%%%%%%%%%%%%%%%%%%%%%%%%%%%%%%%%%%%%%%%%%%%%%%%%%%%%%%
\begin{table*}[ht]
\centering
\caption{The values of proton $g^{p}_{\mathrm{pair}}$ and neutron $g^{n}_{\mathrm{pair}}$ pairs that renormalise the residual interaction and reproduce the respective empirical pairing gaps $\Delta_{p}$ and $\Delta_{n}$. The active model space and the harmonic oscillator parameter, for each isotope, are also presented.}
\label{table1}
%\begin{tabular}{l*{5}{c}llr}
\begin{tabular}{{c|cccccc}}
\hline \hline
Nucleus & model-space & $b$ & $\Delta_p$ & $\Delta_n$ & $g^{p}_{\mathrm{pair}}$ & $g^{n}_{\mathrm{pair}}$\\
\hline
$^{12}\mathrm{C}$ & 8 (no core) & 1.522 & 4.68536 & 4.84431 & 1.12890 & 1.19648 \\
$^{16}\mathrm{O}$ & 8 (no core) & 1.675 & 3.36181 & 3.49040 & 1.06981 &  1.13636 \\
$^{20}\mathrm{Ne}$ & 10 (no core) & 1.727 & 3.81516 & 3.83313 & 1.15397 & 1.27600\\
$^{28}\mathrm{Si}$ & 10 (no core) & 1.809 & 3.03777 & 3.14277 & 1.15568 & 1.23135\\
%$^{32}\mathrm{S}$  & 15 (no core) & 1.843 & 2.03865 & 2.09807 & 0.8837 & 0.95968\\
$^{40}\mathrm{Ar}$ & 15 (no core) & 1.902 & 1.75518 & 1.76002 & 0.94388 & 1.01348\\
$^{48}\mathrm{Ti}$ & 15 (no core) & 1.952 & 1.91109 & 1.55733 & 1.05640 & 0.99890 \\
$^{76}\mathrm{Ge}$ & 15 (no core) & 2.086 & 1.52130 & 1.56935 & 0.95166 & 1.17774 \\
%$^{114}\mathrm{Cd}$ & 18 (core $^{16}\mathrm{O}$) & 2.214 & 1.41232 & 1.35155 & 1.03122 & 0.98703 \\
$^{132}\mathrm{Xe}$ & 15 (core $^{40}\mathrm{Ca}$) & 2.262 & 1.19766 & 1.20823 & 0.98207 & 1.13370 \\
%$^{208}\mathrm{Pb}$ & & & & \\
 \hline \hline
\end{tabular}

\end{table*}
%%%%%%%%%%%%%%%%%%%%%%%%%%%%%%%%%%%%%%%%%%%%%%%%%%%%%%%%%%%%%%%%%

\subsection{Other methods for obtaining the nuclear form factors}
\label{other-methods}
The nuclear form factor, which is the Fourier transform of the nuclear charge density distribution $\rho_{p}(r)$, is defined as
\begin{equation}
F_{Z}(q^2)=\frac{4 \pi}{Z}\int \rho_{p}(r) j_{0}(q r)\, r^2 \, dr \, ,
\label{definition-ff}
\end{equation}
with $j_{0}$ being the zero order spherical Bessel function. Due to the significance of the nuclear form factors in our calculations and for the benefit of the reader we devote a separate discussion to summarise some useful possibilities of obtaining these observables.

i) \emph{Use of available experimental data}\\
For many nuclei and especially for odd-A isotopes, the proton nuclear form factors $F_{Z}(q^{2})$, are computed by means of a model independent analysis (using a Fourier-Bessel expansion model or others) of the electron scattering data for the proton charge density $\rho_{p}(r)$ \cite{deVries} wherever, possible. The absence of similar data for neutron densities, restricts us to assume that  $F_{N}(q^2) = F_{Z}(q^2)$.  In this work, we  consider this method only for the case of the very heavy doubly closed $^{208}\mathrm{Pb}$ nucleus.  
%%%%%%%%%%%%%%%%%%%%%%%%%%%%%%%%%%%%%%%%%%%%%%%%%%%%%%%%%%%%%%%%%%%%%%%%%%%%%%%%%%%%%%%%%%%%%%%%%%%%
%
%%%%%%%%%%%%%%%%%%%%%%%%%%%%%%%%%%%%%%%%%%%%%%%%%%%%%%%%%%%%%%%%%%%
\begin{figure}[ht]
\centering
\includegraphics[width=\textwidth]{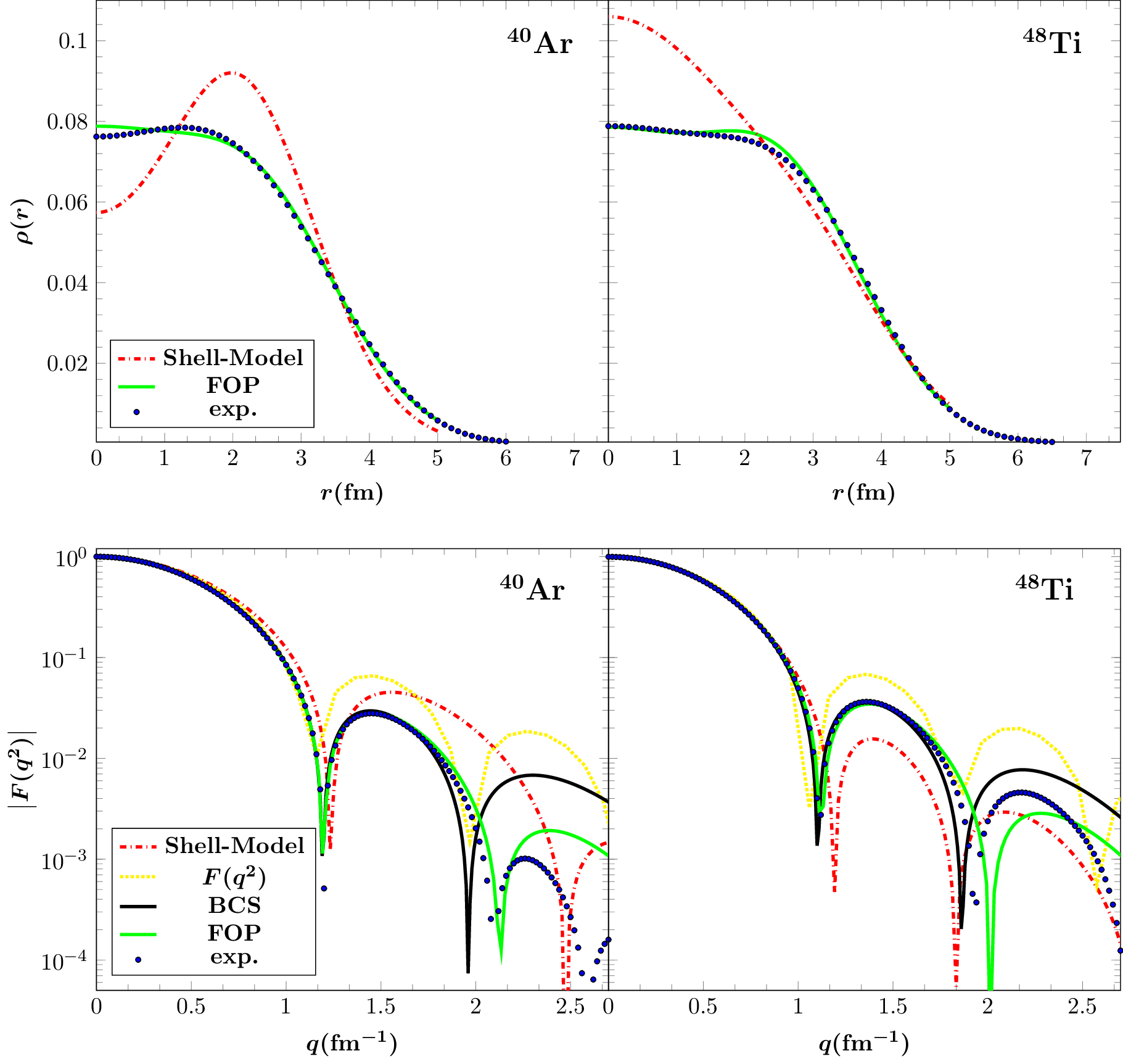}
\caption{ The charge density distribution (\textit{left}) and  the form factor as a function of the momentum transfer (\textit{right}), for the cases of $^{40}$Ar and $^{48}$Ti nuclei. The introduction of fractional occupation probabilities (FOP) of the states provides higher reproducibility of the experimental data, compared to the simple Shell-Model and that of Eq. (\ref{F-bessel}). The BCS nuclear neutron form factor $F_{N}(q^2)$ is also presented and compared.}
\label{fig.2}
\end{figure}
%%%%%%%%%%%%%%%%%%%%%%%%%%%%%%%%%%%%%%%%%%%%%%%%%%%%%%%%%%%%%%%%%%%

ii) \emph{Fractional occupation probabilities in a simple Shell-Model}\\
In Ref. \cite{Kosm_A536} the form factor $F_{Z}(q^2)$,  for h.o. wavefunctions has been written as \cite{Kosm_B215}
\begin{equation}
F_{Z}(q^2)=\frac{1}{Z} e^{-(q\,b)^{2}/4} \Phi \left(q\,b,Z \right), \qquad \Phi\left(q\,b,Z \right)= \sum_{\lambda=0}^{N_{\mathrm{max}}} \theta_{\lambda} (q\,b)^{2 \lambda} \, .
\label{forfac}
\end{equation}
The radial nuclear charge density distribution $\rho_{p}(r)$, entering the definition of Eq. (\ref{definition-ff}), is written in the following compact form \cite{Kosm_B215}
\begin{equation}
\rho_{p}(r)=\frac{1}{\pi^{3/2} b^{3}}e^{-(r/b)^{2}} \, \Pi\left(\frac{r}{b},Z\right), \qquad \Pi\left(\chi,Z\right)=\sum_{\lambda=0}^{N_{\mathrm{max}}} f_{\lambda} \chi^{2 \lambda},
\label{rho(r)}
\end{equation}
where $\chi=r/b$, with $b$ denoting the h.o. size parameter. 
$N_{\mathrm{max}}=(2n+ \ell)_{\mathrm{max}}$ stands for the number of quanta of the highest occupied proton (neutron) level. The coefficients $f_{\lambda}$ are expressed as
\begin{equation}
f_{\lambda}=\sum_{(n,\ell)j} \frac{\pi^{1/2}(2j+1)n! \,C_{n \ell}^{\lambda-\ell} }{2 \Gamma\left( n+\ell+\frac{3}{2} \right)} \, ,
\end{equation}
where $\Gamma(x)$ is the Gamma function.  For the coefficients $\theta_{\lambda}$, $C_{n \ell}^{\lambda-\ell}$ and further information see Ref. \cite{Kosm_B215}.

Up to this point, that the proton occupation probabilities entering Eq. (\ref{definition-ff}) Eq. (\ref{forfac}) have been considered equal to unity for the states below the Fermi surface and zero for those above the Fermi surface. In Ref. \cite{Kosm_A536}, the authors introduced depletion and occupation numbers, to parametrise the partially occupied levels of the states. These parameters satisfy the relation
\begin{equation}
\sum_{\begin{subarray}{c} (n \ell)j\\ \mathrm{all} \end{subarray} } \alpha_{n \ell j}(2j+1) =N_{n}\, .
\label{depletion}
\end{equation} 
 Within this context, the "active" surface nucleons (above or below the Fermi level) have non-zero occupation probability $\alpha_{n \ell j}\neq 0$, smaller than unity, while the "core" levels have occupation probability $\alpha_{n \ell j}=1$. In this paper we extend the work of Ref. \cite{Kosm_A536} where three parameters $\alpha_{1},\alpha_{2},\alpha_{3}$ are used to describe the partial occupation probabilities of the surface orbits. We improve the formalism by introducing more parameters, increasing this way the number of "active" nucleons in the studied nuclear system and come out with higher reproducibility of the  experimental data \cite{deVries}. To this aim, we introduce four parameters $\alpha_{i}, \, \, i=1,2,3,4$ in Eq. (\ref{depletion}).  Then, the assumed "active" single-particle levels are five and Eq. (16) of Ref. \cite{Kosm_A536} becomes
\begin{equation}
\begin{aligned}
\Pi(\chi,Z,\alpha_{i})= & \Pi(\chi,Z_{2}) \frac{\alpha_{1}}{Z_{1}-Z_{2}} + \Pi(\chi,Z_{1}) \left[\frac{\alpha_{2}}{Z_{c}-Z_{1}}-\frac{\alpha_{1}}{Z_{1}-Z_{2}} \right] \\
& + \Pi(\chi,Z_{c}) \left[\frac{Z^{\prime}-Z}{Z^{\prime}-Z_{c}}-\frac{\alpha_{2}}{Z_{c}-Z_{1}}-\frac{\alpha_{3}}{Z^{\prime}-Z_{c}} \right] \\
& + \Pi(\chi,Z^{\prime}) \left[\frac{Z-Z_{c}}{Z^{\prime}-Z_{c}} + \frac{\alpha_{3}}{Z^{\prime}-Z_{c}}- \frac{\alpha_{4}}{Z^{\prime \prime}-Z^{\prime}} \right] \\
& + \Pi(\chi,Z^{\prime \prime}) \left[\frac{\alpha_{4}}{Z^{\prime \prime}-Z^{\prime}} - \frac{\lambda}{Z^{\prime \prime \prime}-Z^{\prime \prime} }\right] + \Pi(\chi,Z^{\prime \prime \prime}) \frac{\lambda}{Z^{\prime \prime \prime}-Z^{\prime \prime}}\, ,
\end{aligned}
\label{fractional}
\end{equation}
with $\lambda=\alpha_{1}+\alpha_{2}-\alpha_{3}-\alpha_{4}$.
By substituting the polynomial $\Pi(\chi,Z)$ of Eq. (\ref{rho(r)}) with that of the latter expression and using the experimental data \cite{deVries}, we  we fit the parameters $\alpha_{i}$ (and similarly for the form factor of Eq. (\ref{forfac})). As an example, for the $^{40}$Ar isotope we have, $Z_{2}=10,\, Z_{1}=12, \, Z=Z_{c}=18, \, Z^{\prime}=20, \,   Z^{\prime \prime}=22,\,  Z^{\prime \prime \prime}=30$. The resulting fractional occupation probabilities that fit the experimental charge density distribution are $\alpha_{1} = 0.85$, $\alpha_{2} = 1.25$, $\alpha_{3} = 0.85$, $\alpha_{4} = 0.75$. Similarly for the $^{48}$Ti nucleus, we have $Z_{2}=18,\, Z_{1}=20, \, Z=Z_{c}=22, \, Z^{\prime}=30, \,   Z^{\prime \prime}=34,\,  Z^{\prime \prime \prime}=40$ and  the fitting parameters are $\alpha_{1} = 1.0$, $\alpha_{2} = 1.5$, $\alpha_{3} = 0.35$, $\alpha_{4} = 0.1$. In Fig.  \ref{fig.2} the prediction of the method is compared with that of the simple shell-model and the experimental data. We note that in the momentum transfer range of our interest (i.e. $q<2 \, \mathrm{fm}^{-1}$) the form factor has excellent behaviour. We however mention that even though the FOP method presents very high reproducibility of the experimental data, it is not always applicable, e.g. for deformed nuclei (where BCS appears to be still successful).

iii) \emph{Use of effective expressions for the nuclear form factors}\\
We finally discuss one of the most accurate effective methods for calculating the nuclear form factor by  Ref. \cite{Engel} 
\begin{equation}
F(q^2)=\frac{3 j_{1}(qR_{0})}{qR_{0}}\exp \left[-\frac{1}{2} (q s)^2 \right]\, ,
\label{F-bessel}
\end{equation}
where $j_{1}(x)$ is the known first-order Spherical-Bessel function and  $R_{0}^{2}=R^{2}-5 s^{2}$, with $R$ and $s$ being the radius and surface thickness parameters of the nucleus respectively. The radius parameter is usually given from the semi-empirical form $R=1.2 A^{1/3}\,\mathrm{fm}$ while $s$ is of the order of $0.5 \,\mathrm{fm}$ (see Ref. \cite{deVries}). 

It is worth noting that, by inserting the form factors $F_{Z(N)}$ obtained as described above in Eq. (\ref{SM-ME}), the resulting cross sections have a rather high confidence level. In the next part of the paper the results show that the momentum dependence of the nuclear form factors becomes crucial, especially for intermediate and high energies. In some cases, differences of even an order of magnitude may occur as compared to the calculations neglecting the momentum dependence of the nuclear form factors. 

%%%%%%%%%%%%%%%%%%%%%%%%%%%%%%%%%%%%%%%%%%%%%%%%%%%%%%%%%%%%%%%%%
%%%%%%%%%%%%%%%%%%%%%%%%%%%%%%%%%%%%%%%%%%%%%%%%%%%%%%%%%%%%%%%%%
\section{Results and discussion}
\label{results}
%%%%%%%%%%%%%%%%%%%%%%%%%%%%%%%%%%%%%%%%%%%%%%%%%%%%%%%%%%%%%%%%%
\subsection{Integrated coherent $\nu$-nucleus cross sections}
\begin{figure}[ht]
\begin{center}
\includegraphics[width=0.5\textwidth]{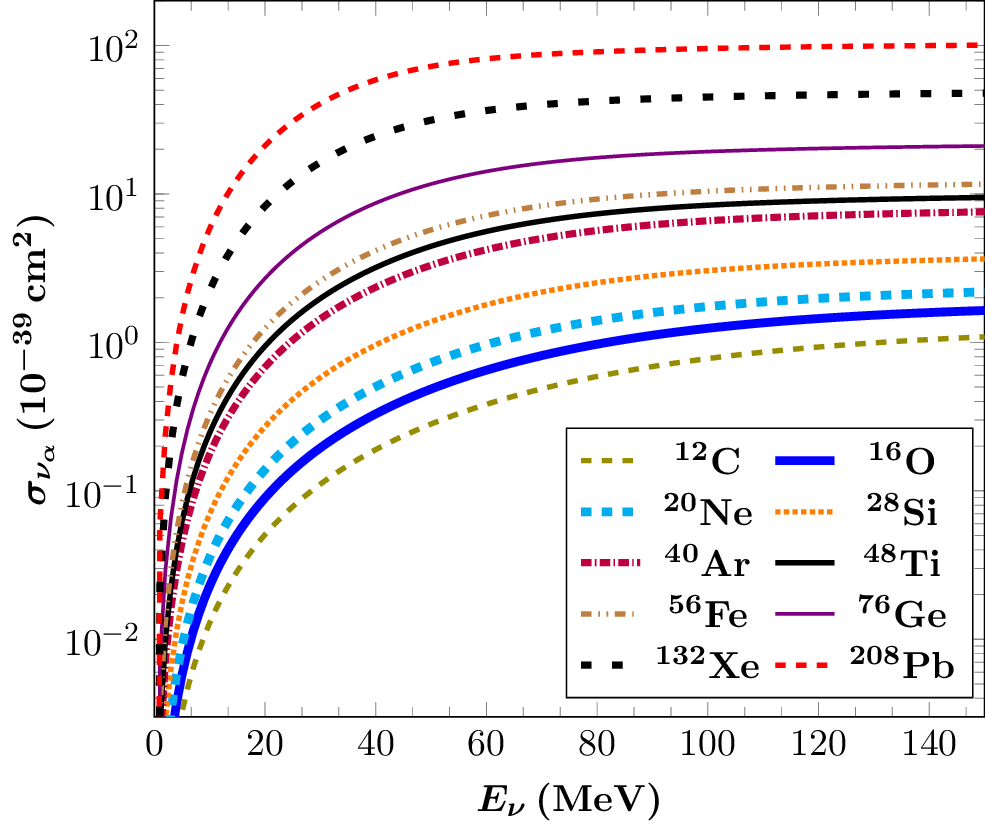}
\end{center}
\caption{Total coherent cross sections  $\sigma_{\nu_{\alpha}(\tilde{\nu}_{\alpha})}(E_\nu)$ in units $10^{-39} \mathrm{cm^2}$ for a set of nuclei as a function of the incoming 
neutrino energy $E_\nu$, for the SM neutrino processes
$\nu_{\alpha}(\tilde{\nu}_{\alpha}) + (A,Z) \rightarrow  \nu_{\alpha}(\tilde{\nu}_{\alpha}) + (A,Z)$.} 
\label{fig.3} 
\end{figure}

%%%%%%%%%%%%%%%%%%%%%%%%%%%%%%%%%%%%%%%%%%%%%%%%%%%%%%%%%%%%%%%%%
The next phase of our calculational procedure is related to the total coherent $\nu$-nucleus cross sections,  obtained through numerical integration of Eq. (\ref{SM_dcostheta}) over angles [or Eq. (\ref{SM_dT}) over $T_N$] as
\begin{equation}
\sigma_{\nu_{\alpha}}(E_\nu) = \int\frac{d\sigma_{\nu_{\alpha}}}{d\cos \theta}(\theta , 
E_\nu) \,\, d\cos \theta  \, . 
\end{equation} 
 The results for the Standard Model cross sections, for a set of different promising targets throughout the periodic table, are presented in Fig.  \ref{fig.3}.  As can be seen, the present nuclear structure calculations indicate that between light and heavy nuclear systems, the cross sections may differ by even two orders  of magnitude (or more) as a consequence of the dependence on the nuclear parameters (i.e. mass, form factors, etc.). We also see that for heavier nuclei the cross sections flatten more quickly (at lower neutrino energies) compared to that of lighter nuclear isotopes. The latter conclusion originates mainly from the fact that, for heavy nuclei the suppression of the cross sections due to the nuclear form factors  becomes more significant. Thus, for heavy material the nuclear effects become important even at low energies. Such original cross section results are helpful for the simulations of the Standard  and non-standard Model signals of $\nu$-detection experiments (see below).
%
%%%%%%%%%%%%%%%%%%%%%%%%%%%%%%%%%%%%%%%%%%%%%%%%%%%%%%%%%%%%%%%%%%%
%%%%%%%%%%%%%%%%%%%%%%%%%%%%%%%%%%%%%%%%%%%%%%%%%%%%%%%%%%%%%%%%%%%
%
%%%%%%%%%%%%%%%%%%%%%%%%%%%%%%%%%%%%%%%%%%%%%%%%%%%%%%%%%%%%%%%%%

\subsection{Supernova neutrino simulations}

As discussed previously, our present calculations may also be useful for ongoing and future neutrino experiments related to supernova (SN) neutrino detection, since as it is known, the neutrinos emitted in SN explosions transfer the maximum part of the total the energy released. Then, the total neutrino flux, $\Phi(E_\nu)$, arriving at a terrestrial detector as a function of the SN neutrino energy $E_\nu$, the number of emitted (anti)neutrinos $N_{\nu_{\alpha}}$ at a 
distance $d$ from the source (here we consider $d = 10 \, \mathrm{kpc}$), reads \cite{Horowitz,Biassoni}
\begin{equation}
\Phi(E_\nu) = \sum_{\nu_{\alpha}} \Phi_{\nu_{\alpha}}  \eta_{\nu_{\alpha}}^{\mathrm{SN}} (E_\nu)=\sum_{\nu_{\alpha}} \frac{N_{\nu_{\alpha}}}{4\pi \, d^ 2}\, 
\eta_{\nu_{\alpha}}^{\mathrm{SN}} (E_\nu)\, ,
\end{equation}
($\alpha = e, \mu, \tau$) where  
$\eta_{\nu_{\alpha}}^{\mathrm{SN}}$ denotes the energy distribution 
of the (anti)neutrino flavour $\alpha$. 

The emitted SN-neutrino energy spectra $\eta_{\nu_{\alpha}}^{\mathrm{SN}} (E_\nu)$ 
may be parametrised by Maxwell-Boltzmann distributions that depend only on the temperature $T_{\nu_{\alpha}}$ of the 
(anti)neutrino flavour $\nu_\alpha$ or $\tilde{\nu}_\alpha$ (the chemical potential is ignored) we have
\begin{equation}
\eta_{\nu_\alpha}^{\mathrm{SN}}(E_\nu) = \frac{E_\nu^2}{2 T_{\nu_\alpha}^3} e^{-E_\nu/T_{\nu_\alpha}} \, ,
\label{eta-SN}
\end{equation} 
($T_{\nu_e}=3.5 \mathrm{MeV}$, $T_{\tilde{\nu}_e} = 5.0 \mathrm{MeV}$, $T_{\nu_x,\tilde{\nu}_x} = 8.0\mathrm{MeV}$, 
$x = \mu, \tau$ \cite{Giom-Vergados}). For each flavour, the total number of emitted neutrinos $N_{\nu_{\alpha}}$ is obtained from the mean neutrino energy \cite{PLB}
\begin{equation}
\langle E_{\nu_\alpha}\rangle = 3 T_{\nu_{\alpha}}
\end{equation} 
and the total energy released from a SN explosion, $U=3 \times 10^{53} \mathrm{erg}$ \cite{Hirata-Bionta}.
\subsection{Laboratory neutrino simulations}
\label{lab_neutrinos}
 
The spallation neutron source (SNS) at Oak Ridge National Lab \cite{SNS} produces neutrons by firing a pulsed proton beam at a liquid mercury target \cite{Amanik-McLaughin}. The main aim of the COHERENT proposal \cite{coherent1,coherent2} (or of other similar conceptual \cite{CLEAR,BNB}) concerns with possible detection of neutrino-nucleus coherent scattering events at the SNS. Our simulations here, are mainly motivated by previous studies \cite{Scholberg,Collar,sterile,Brice} and the hope to provide our accurate nuclear structure calculations.

In stopped pion-muon sources, neutrinos are produced by the pion decay chain. Pion decay at rest $\pi^+ \rightarrow \mu^{+} \nu_{\mu}, $ ($\tau=26 \, \mathrm{ns}$) produces monochromatic muon neutrinos $\nu_{\mu}$ at 29.9 MeV, followed by electron neutrinos $\nu_e$ and muon antineutrinos $\tilde{\nu}_{\mu}$ that are produced by the muon-decay $\mu^{+} \rightarrow \nu_{e} e^{+} \tilde{\nu}_{\mu}$ ($\tau=2.2 \, \mathrm{\mu s}$) \cite{Avignone-Efremenko,Efremenko-2009}. For pulsed beams in time-scales narrower than $\mathrm{\mu s}$, $\nu_{e}$'s and $\tilde{\nu}_{\mu}$'s will be delayed with the beam while $\nu_{\mu}$'s will be prompt with the beam \cite{Scholberg}. The emitted $\nu_{e}$ and ${\tilde{\nu}_{\mu}}$ neutrino spectra are described by the high precision normalized distributions, known as the Michel spectrum \cite{Louis}
 \begin{equation}
 \begin{aligned}
\eta_{\nu_{e}}^{lab.}=& 96 E_{\nu}^{2}M_{\mu}^{-4} \left( M_{\mu}-2E_{\nu}\right)\, ,\\
\eta_{\tilde{\nu}_{\mu}}^{lab.}=& 16 E_{\nu}^{2}M_{\mu}^{-4} \left( 3 M_{\mu}-4E_{\nu}\right)\, ,
\end{aligned}
\label{labor-nu}
\end{equation}
($M_{\mu}=105.6 \, \mathrm{MeV}$ is the muon rest mass).  
The maximum neutrino energy in the latter distributions is $E_{\nu}^{\text{max}}=M_{\mu}/2 =52.8 \, \mathrm{MeV}$
(see e.g. \cite{pion-DAR-nu}). 

The spallation neutron source (SNS) at Oak Ridge National Lab is currently the most powerful facility to detect for a first time neutrino-nucleus coherent scattering events, since it provides exceptionally intense fluxes $\Phi_{\nu_{\alpha}} = 2.5 \times 10^{7} \, \nu \mathrm{s^{-1} cm^{2}}$ at 20 m and $\Phi_{\nu_{\alpha}} = 6.3 \times 10^{6} \, \nu \mathrm{s^{-1} cm^{2}}$ at 40 m from the source \cite{Avignone-Efremenko,Efremenko-2009}. The simulated laboratory neutrino signals $\sigma_{\nu, lab.}^{sign}$ coming out of our calculations for the adopted nuclear targets are discussed below.

\subsection{Simulated neutrino signals}
\begin{figure}[ht]
\begin{center}
\includegraphics[width=\textwidth]{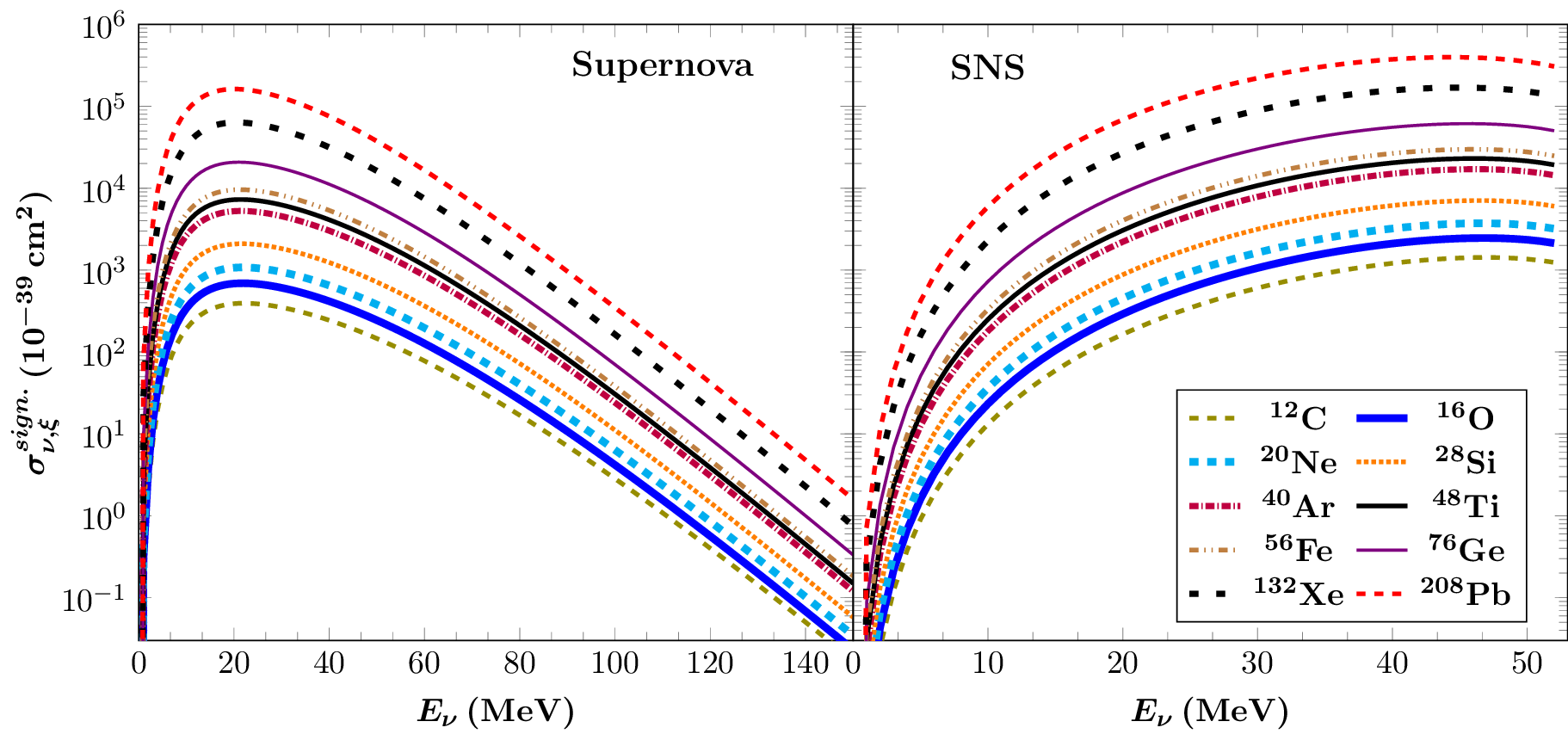}
\end{center}
\caption{The signal cross sections that represent the expected signal to be recorded on a terrestrial nuclear $\nu$-detector, (\emph{left}) for supernova neutrinos ($\xi=\mathrm{SN}$), evaluated with Maxwell-Boltzmann distributions at $d = 10 \, \mathrm{kpc}$, and (\emph{right}) for SNS neutrinos ($\xi=lab.$), at 20 m from the source. For the case of SNS neutrinos the figure takes into account only the delayed beam, evaluated with the generic flux of $\Phi_{\nu_{\alpha}} \sim 10^7\, \nu \mathrm{s^{-1} cm^{-2}}$. Different nuclear detectors have been studied.}\label{fig.4}
\end{figure}

By weighting the integrated cross section $\sigma_{\nu_\alpha}(E_\nu)$ with the neutrino distributions of Eq. (\ref{eta-SN}), for SN neutrinos, or Eq. (\ref{labor-nu}), for laboratory neutrinos, the total signal produced on a terrestrial detector is described by \cite{vtsak-tsk-1}
\begin{equation}
\sigma^{sign}_{\nu, \xi}(E_\nu)=\sum \limits_{ \nu_{\alpha}}\sigma_{\nu_\alpha}(E_\nu)\,\eta_{\nu_\alpha}^{\xi}(E_\nu), \quad \xi=\mathrm{SN},lab.
\label{eq.signal}
\end{equation}
%
%%%%%%%%%%%%%%%%%%%%%%%%%%%%%%%%%%%%%%%%%%%%%%%%%%%%%%%%%%%%%%%%% 
The resulting signals, $\sigma^{sign}_{\nu,\xi}(E_\nu)$, obtained by inserting in Eq. (\ref{eq.signal}) the  cross sections $\sigma_{\nu_{\alpha}}$ of Fig. \ref{fig.3}, are plotted 
in Fig. \ref{fig.4}.
%
%%%%%%%%%%%%%%%%%%%%%%%%%%%%%%%%%%%%%%%%%%%%%%%%%%%%%%%%%%%%%%%%%%%
%%%%%%%%%%%%%%%%%%%%%%%%%%%%%%%%%%%%%%%%%%%%%%%%%%%%%%%%%%%%%%%%%%% 
%

In our previous work \cite{PLB} it was shown that the simulated cross sections reflect the characteristics of the incident neutrino spectrum of the specific neutrino flavour $\alpha$ and therefore, such a simulated signal is characterised by its own position of the maximum peak and width of the distribution $\eta_{\nu_\alpha}^{\mathrm{SN}}$. We, however, remind that within the framework of the SM, coherent neutrino scattering is a flavour blind and a particle-antiparticle blind process. For this particular case our results are shown in Fig. \ref{fig.4} for supernova and laboratory (SNS) neutrinos.

In neutrino simulations, another useful quantity is the flux averaged cross section 
\cite{Kosm-Oset} which in our notation is written as
\begin{equation}
\langle\sigma_{\nu}\rangle_{\xi} = \sum \limits_{ \nu_{\alpha}} \int\sigma_{\nu_\alpha}(E_\nu)\,
\eta_{\nu_\alpha}^{\xi}(E_\nu) \, dE_\nu \, .
\end{equation}
%%%%%%%%%%%%%%%%%%%%%%%%%%%%%%%%%%%%%%%%%%%%%%%%%%%%%%%%%%%%%%%%%
\begin{table*}[ht]
\centering
%\begin{tabular}{l*{5}{c}llr}
\caption{Flux averaged cross sections $\langle\sigma_{\nu}\rangle_{\xi}$ in units $10^{-40} \mathrm{cm^{2}}$ for the adopted supernova ($d=10 \, \mathrm{kpc}$) and laboratory (delayed flux only) neutrino spectra. For the case of SNS neutrinos, we adopt the generic flux, i.e. $\Phi_{\nu_{\alpha}} \sim 10^{7} \, \nu \mathrm{s^{-1} cm^{2}}$ at 20 m for all nuclear targets.}
\begin{tabular}{{c|cccccccccccc}}
\hline \hline
Nucleus & $^{12}\mathrm{C}$ & $^{16}\mathrm{O}$ &  $^{20}\mathrm{Ne}$ &   $^{28}\mathrm{Si}$ & $^{40}\mathrm{Ar}$ & $^{48}\mathrm{Ti}$ & $^{56}\mathrm{Fe}$ & $^{76}\mathrm{Ge}$ & $^{132}\mathrm{Xe}$ & $^{208}\mathrm{Pb}$\\
\hline
$\langle\sigma_{\nu}\rangle_{\mathrm{SN}}$ & 1.46 & 2.51 & 3.91 & 7.52 & 18.59 & 25.43 & 33.29 & 70.63 & 207.56 & 514.93\\
$\langle\sigma_{\nu}\rangle_{lab.}$ & 3.07 & 5.33 & 8.13 & 15.52 & 37.91 & 51.50 & 67.02 & 139.83 & 395.59 & 949.50\\
 \hline \hline
\end{tabular}

\label{table2}
\end{table*}
%%%%%%%%%%%%%%%%%%%%%%%%%%%%%%%%%%%%%%%%%%%%%%%%%%%%%%%%%%%%%%%%%
The results for $\langle\sigma_{\nu}\rangle_{\xi}$, obtained by using the angle-integrated cross sections 
of Fig. \ref{fig.3} are listed in Table \ref{table2} for both neutrino sources. 

%%%%%%%%%%%%%%%%%%%%%%%%%%%%%%%%%%%%%%%%%%%%%%%%%%%%%%%%%%%%%%%%%%%%%%
\subsection{Differential and total event rates}
\begin{figure}[ht]
\centering
\includegraphics[width=\textwidth]{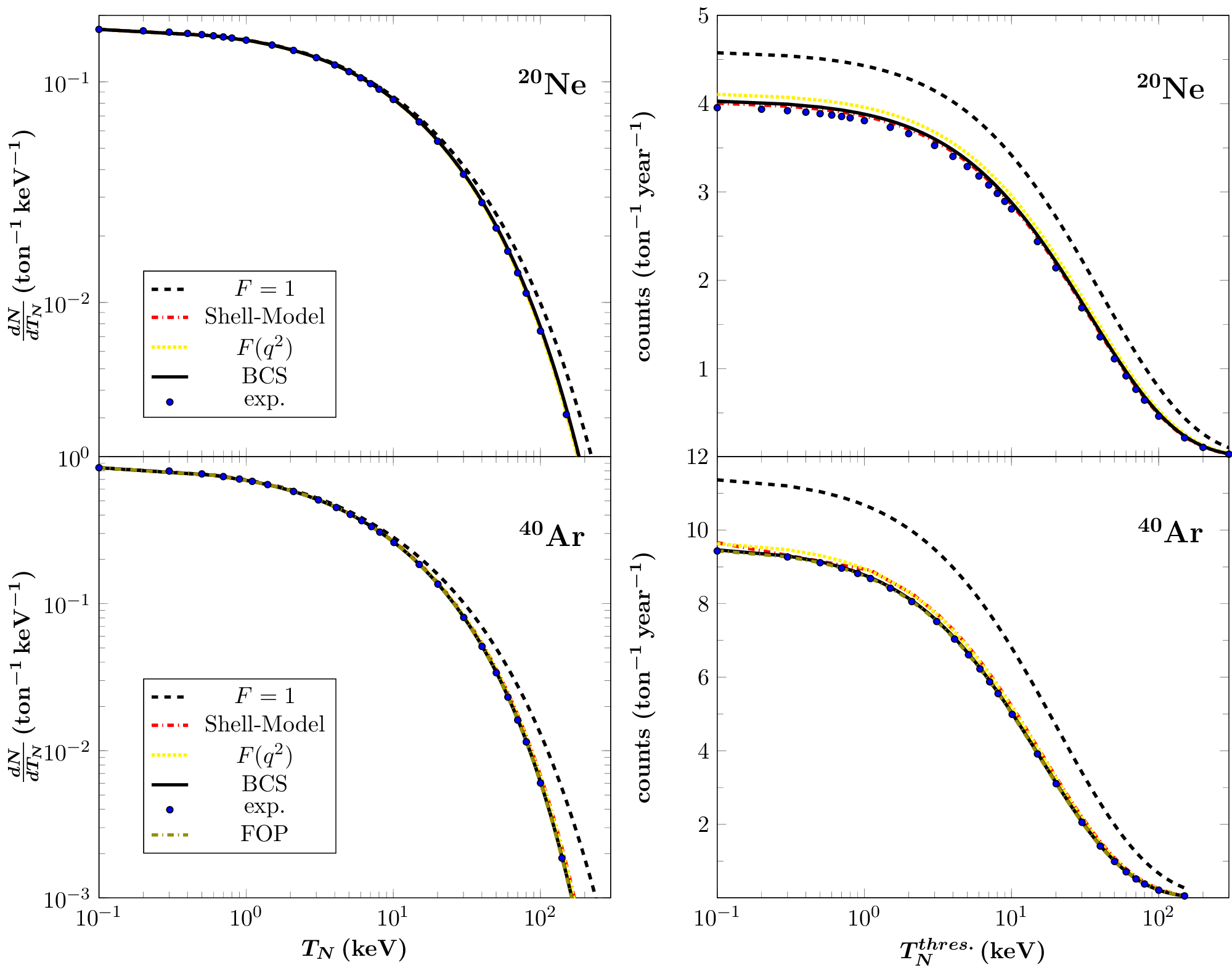}
\caption{Yield in events  (\textit{left}) and total number of events over  nuclear recoil threshold $T_{N}^{thres.}$ (\textit{right}), for supernova neutrinos at $d = 10 \, \mathrm{kpc}$. Here, 1 ton of perfectly efficient $^{20}$Ne and $^{40}$Ar detectors have been considered and also possible neutrino oscillation in propagation effects are neglected. For heavier nuclear targets the differences become rather significant. In this figure, $F(q^2)$ stands for Eq. (\ref{F-bessel}) and FOP for the method of fractional occupation probabilities of the states. For more details see the text.}\label{fig.5}
\end{figure}

\begin{figure}[ht]
\centering
\includegraphics[width=\textwidth]{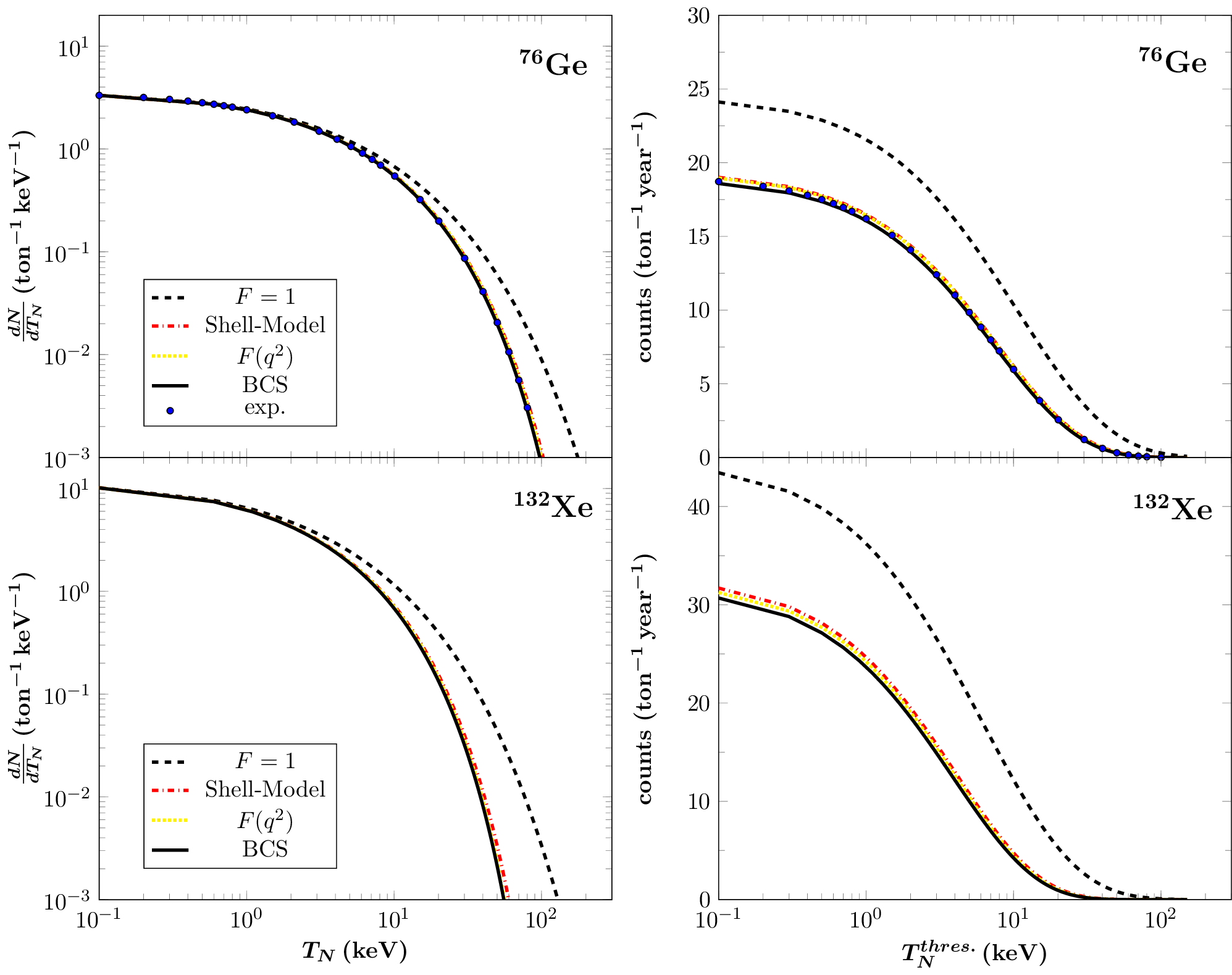}
\caption{Same as Fig. \ref{fig.5} but for $^{76}$Ge and $^{132}$Xe.}\label{fig.6}
\end{figure}

From experimental physics perspectives, predictions for the differential 
event rate, $Y_{\nu_{\alpha}}$, of a $\nu$-detector are crucial \cite{Horowitz}. The usual expression 
for computing the yield in events is based on the neutrino flux, and  is defined as \cite{Biassoni}
\begin{equation}
Y_{\nu_\alpha}(T_N) =\frac{dN}{T_N}= K \sum \limits_{\nu_{\alpha}} \Phi_{\nu_{\alpha}} \int \eta_{\nu_\alpha}^{\xi} \, dE_{\nu} \int \frac{d\sigma_{\nu_\alpha}}{d\cos\theta} \,   
\delta\left(T_N - \frac{q^2}{2 M}\right) \, d\cos \theta \, , 
\label{diff.rate}
\end{equation}
where $K=N_{targ.} t_{tot.}$ accounts for the total number of nuclei (atoms) in the detector material $N_{targ.}$ times the total time of exposure $t_{tot}$. Using the latter equation, one concludes that, the lower the energy recoil, the larger the potentially detected number of events (see Fig. \ref{fig.5} and Fig. \ref{fig.6}). In principle, in order to maximize the potential detection of a rare event process like the $\nu$-nucleus scattering, detector materials with very low energy-recoil threshold and low-background are required. 
%%%%%%%%%%%%%%%%%%%%%%%%%%%%%%%%%%%%%%%%%%%%%%%%%%%%%%%%%%%%%%%%%%%
%
%
%%%%%%%%%%%%%%%%%%%%%%%%%%%%%%%%%%%%%%%%%%%%%%%%%%%%%%%%%%%%%%%%%%%
%%%%%%%%%%%%%%%%%%%%%%%%%%%%%%%%%%%%%%%%%%%%%%%%%%%%%%%%%%%%%%%%%%%
%
In the last stage of our study we make predictions for the total number of coherent scattering events, the most important quantity,  both from theoretical and experimental perspectives. To this purpose, we evaluate the number of expected counts, for the studied detector materials, by performing numerical integration of Eq. (\ref{diff.rate}) over the nuclear recoil threshold $T_N^{thres.}$ (see Table \ref{table3}). 

As has been discussed previously \cite{Horowitz,Anderson}, SN neutrino detection might become possible by the massive dark matter detectors \cite{CLEAN} which have very good energy resolution and low threshold capabilities \cite{Biassoni}. These experiments are designed (or planned) to search for WIMPs \cite{Kosmas-dark_matter,Kosmas-cdm} and/or other rare events such as the neutrinoless double beta decay. The latter, use heavy nuclei as nuclear detectors, e.g. Ge (GERDA \cite{GERDA} and SuperCDMS \cite{SuperCDMS} experiments). In addition we report that SN neutrino events can be potentially detected by experiments using noble gases like Ne (CLEAN detector \cite{CLEAN}), Ar (WARP programme \cite{WARP}) and Xe (XENON 100 Collaboration \cite{XENON-100}).
%%%%%%%%%%%%%%%%%%%%%%%%%%%%%%%%%%%%%%%%%%%%%%%%%%%%%%%%%%%%%%%%%
\begin{table*}[ht]
\centering
%\begin{tabular}{l*{5}{c}llr}
\caption{Total number of events per ton of the target materials for a supernova at a distance of 10 kpc. We assume various energy thresholds 5, 10, 25 or 50 keV. Our present results are in excellent agreement with those of Refs. \cite{Horowitz,Biassoni}.}
\begin{tabular}{{c|cccccc}}
\hline \hline
Nucleus & $T_{N}$ & $T_{N}>5$ keV &  $T_{N}>10$ keV &  $T_{N}>25$ keV &  $T_{N}>50$ keV \\
\hline
$^{12}\mathrm{C}$ & 2.52 & 2.25 & 2.05 & 1.60 & 1.14 \\
$^{16}\mathrm{O}$ & 3.29 & 2.84 & 2.51 & 1.83 & 1.19 \\
$^{20}\mathrm{Ne}$ & 4.03 & 3.35 & 2.87 & 1.96 & 1.16\\
% $^{27}\mathrm{Al}$ & 5.78 & 4.53 & 3.71 & 2.27 & 1.17 \\
%$^{32}\mathrm{S}$  & 6.23 & 4.68 & 3.72 & 2.12 & 0.99 \\
$^{40}\mathrm{Ar}$ & 9.46 & 6.63 & 5.01 & 2.53 & 1.00 \\
$^{48}\mathrm{Ti}$ & 10.73 & 7.04 & 5.06 & 2.27 & 0.76\\
$^{56}\mathrm{Fe}$ & 12.00 & 7.36 & 5.04 & 2.01 & 0.57\\
$^{76}\mathrm{Ge}$ & 18.58 & 9.61 & 5.82 & 1.70 &  0.30 \\
%$^{114}\mathrm{Cd}$ & 26.08 & 9.79 & 4.68 & 0.73 & 0.05\\
$^{132}\mathrm{Xe}$ & 30.68 & 9.84 & 4.16 & 0.46 & 0.01\\
$^{208}\mathrm{Pb}$ & 46.93 & 7.86 & 1.95 & 0.03 & $<10^{-3}$\\
 \hline \hline
\end{tabular}

\label{table3}
\end{table*}
%%%%%%%%%%%%%%%%%%%%%%%%%%%%%%%%%%%%%%%%%%%%%%%%%%%%%%%%%%%%%%%%%

As mentioned in section \ref{nucl-methods}, in order to test our nuclear calculations we have also employed other nuclear methods. To this purpose, we have compared our original results evaluated with the BCS method with those obtained as discussed in subsection \ref{other-methods} and concluded that for the case of the coherent channel all available nuclear methods are in good agreement, but their results differ significantly from those obtained assuming $F_Z(q^2)=F_N(q^2)=1$ (see Fig. \ref{fig.5} and Fig. \ref{fig.6}). We stress however, that, since the cross section is mostly sensitive to the neutron distribution of the target nucleus, the most accurate method (at low and intermediate energies) is the BCS method which provides realistic proton as well as neutron form factors. All other methods employed here consider only the proton distribution and assume $F_{Z}(q^2) = F_{N}(q^2)$, which especially for heavy nuclei, is a rather crude approximation. We remark, however, that the aforementioned nuclear methods offer reliable results on the differential and total event rates for low energies (see Fig. \ref{fig.5} and Fig. \ref{fig.6}), but in order to correctly estimate the neutron form factor, methods like the BSC are probably more appropriate.
%%%%%%%%%%%%%%%%%%%%%%%%%%%%%%%%%%%%%%%%%%%%%%%%%%%%%%%%%%%%%%%%%%% 

\begin{figure}[ht]
\begin{center}
\includegraphics[width=0.5\textwidth]{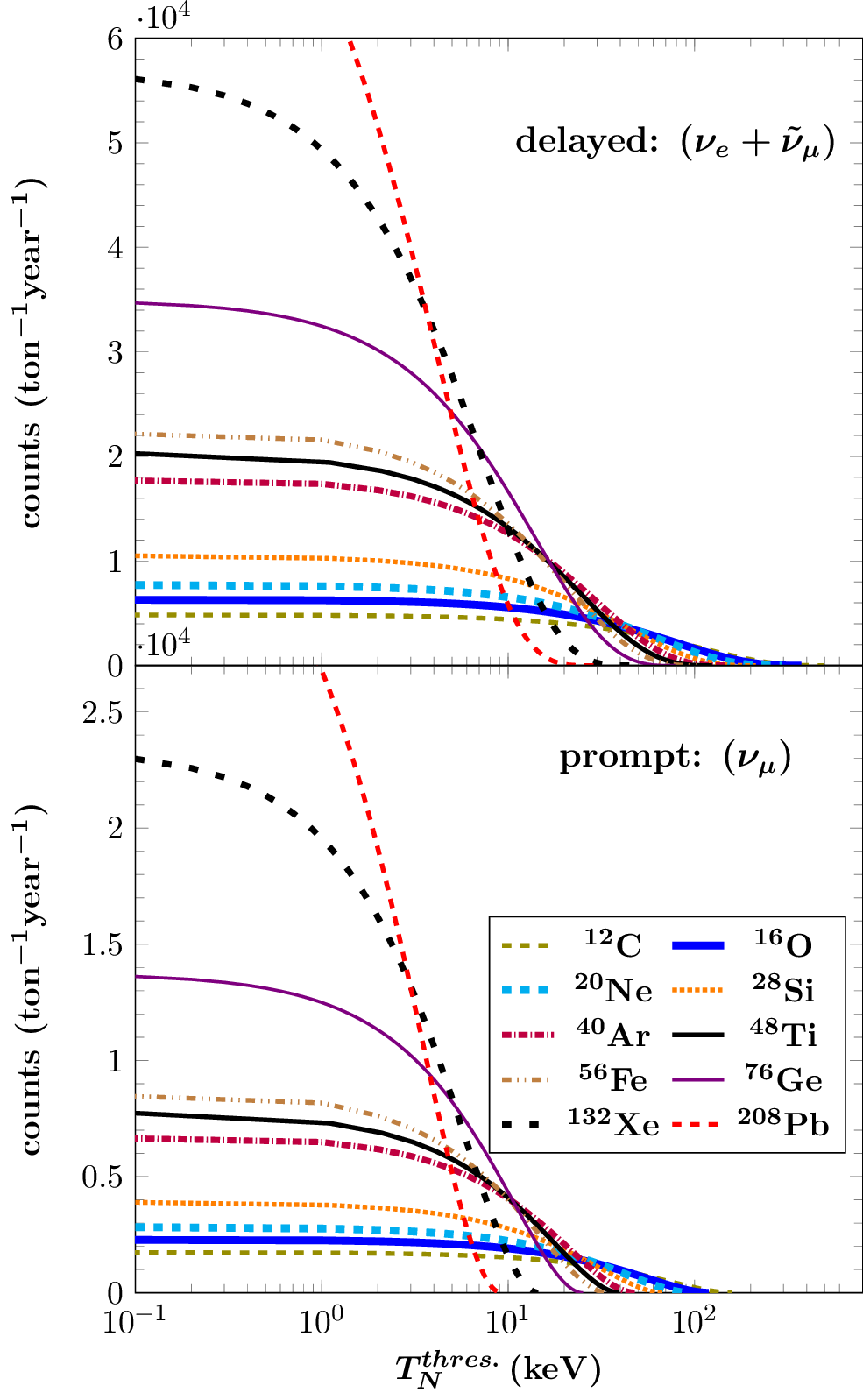}
\end{center}
\caption{Total number of expected events over nuclear recoil threshold for 1 ton of various nuclear targets at 20 m from the source ($\Phi_{\nu_{\alpha}} \sim 10^7\, \nu \mathrm{s^{-1} cm^{-2}} $). The upper (lower) panel assumes the delayed (prompt) flux of laboratory stopped-pion neutrino sources. This figure assumes a perfectly efficient detector and negligible neutrino oscillation effects.}\label{fig.7}
\end{figure}

Our present nuclear structure calculations for laboratory (SNS) neutrinos \cite{SNS} (see Fig. \ref{fig.7}), are in good agreement with previous results \cite{Scholberg}. They imply that a comparably large number of coherent neutrino scattering events is expected to be measured by using LNe, LAr, LXe, Ge and CsI[Na] materials adopted by the COHERENT Collaboration \cite{coherent1,coherent2}. The predictions of the BCS method for these nuclei are illustrated in Fig. \ref{fig.7} and compared with those of other promising nuclear targets. Because the   neutrino flux produced at the SNS is very high, (of the order of $\Phi_{\nu_{\alpha}} \sim 10^{7} \nu \, \mathrm{s}^{-1} \mathrm{cm}^{-2}$ per flavour  at 20 m from the source \cite{Avignone-Efremenko}), even kg-scale experiments expect to measure neutrino-nucleus coherent scattering events at significantly higher rates than those of supernova neutrinos.

It is worth noting that, the choice of the target nucleus plays also a crucial role, since a light nuclear target may yield almost constant number of events throughout the energy range, but small number of counts. On the other hand, a heavy nuclear 
target provides more counts, but yields low-energy recoil making the detection more difficult. This leads to the conclusion that the most appropriate choice for a nuclear detector might be a combination of light and heavy nuclear isotopes, like the scintillation detectors discussed in Ref. \cite{Biassoni}. 
%
%%%%%%%%%%%%%%%%%%%%%%%%%%%%%%%%%%%%%%%%%%%%%%%%%%%%%%%%%%%%%%%%%%%
%%%%%%%%%%%%%%%%%%%%%%%%%%%%%%%%%%%%%%%%%%%%%%%%%%%%%%%%%%%%%%%%%%% 
%

\subsection{Non-standard neutrino interactions at the COHERENT detector}
\label{section-COHERENT}
The multi-target approach of the COHERENT experiment \cite{coherent1,coherent2} aiming on neutrino detection can also explore non-standard physics issues such as NSI \cite{Barranco,PLB},  neutrino magnetic moment \cite{Healey} and sterile neutrino \cite{sterile}. In this subsection we find it interesting to evaluate the  non-standard neutrino-nucleus events that could be potentially detected by this experiment in each of the proposed nuclear targets. The high intensity SNS  neutrino beams \cite{SNS} and the two promising $\nu$-detectors, liquid $^{20} \mathrm{Ne}$ (391 kg)  and liquid $^{40}\mathrm{Ar}$ (456 kg) \cite{sterile}, firstly proposed by the CLEAR \cite{CLEAR} and CLEAN \cite{CLEAN} designs (located at distance 20 m from the source), constitute excellent probes to search for the exotic $\nu$-reactions. Other possibilities \cite{coherent1,coherent2} include medium and heavy weight targets like  $^{76} \mathrm{Ge}$ (100 kg) inspired by the dark matter SuperCDMS \cite{SuperCDMS} detector (located at 20 m) and $^{132} \mathrm{Xe}$ (100 kg located at 40 m).

%
%%%%%%%%%%%%%%%%%%%%%%%%%%%%%%%%%%%%%%%%%%%%%%%%%%%%%%%%%%%%%%%%%%%
\begin{figure}[ht]
\begin{center}
\includegraphics[width=1.0\textwidth]{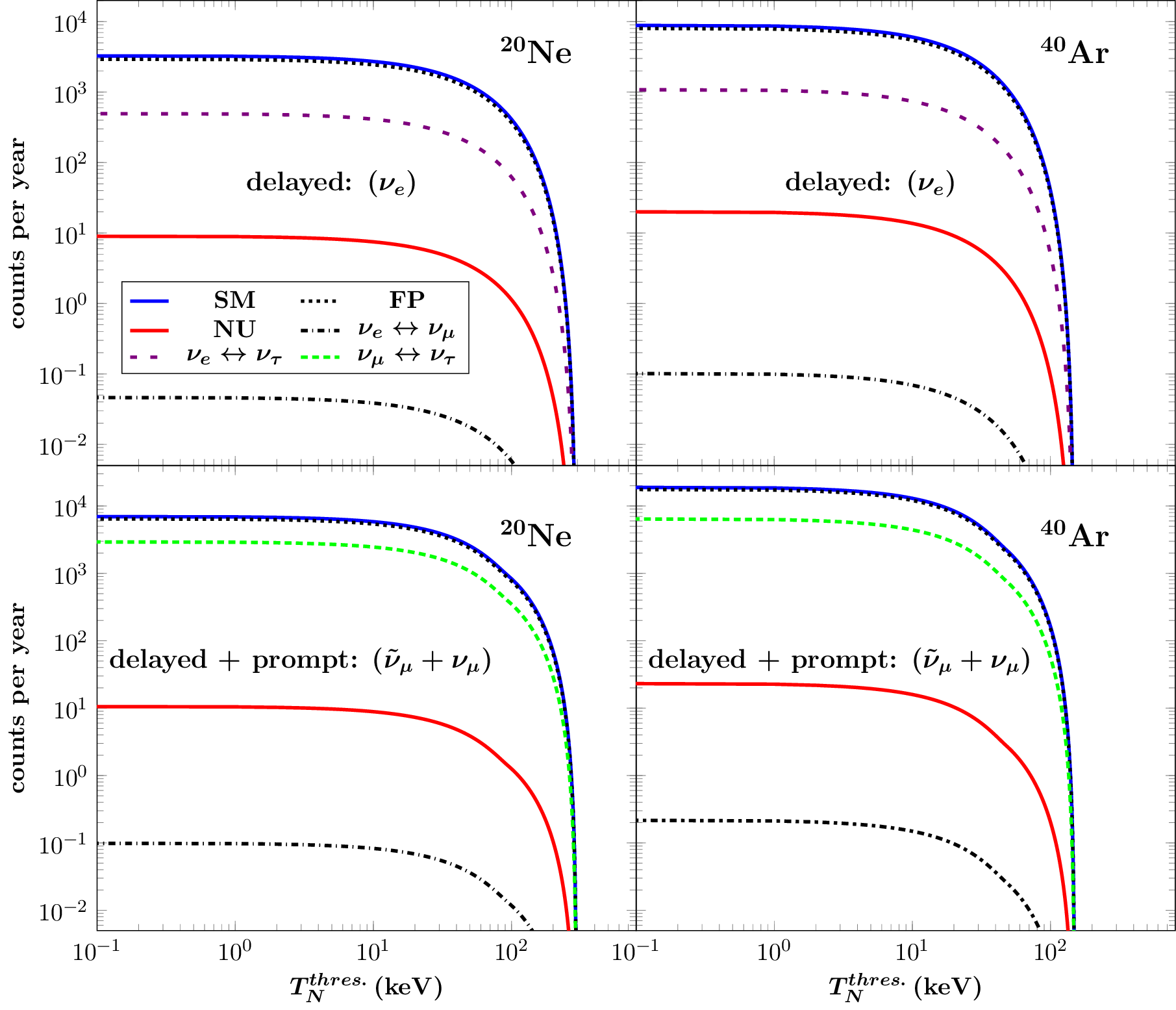}
\end{center}
\caption{The expected non-standard neutrino scattering events over the recoil energy threshold at the COHERENT detector, filled with (\emph{left}) 391 kg of liquid $^{20}\mathrm{Ne}$ and (\emph{right}) 456 kg of liquid $^{40}\mathrm{Ar}$, both located at a distance of 20 m ($\Phi_{\nu_{\alpha}}=2.5 \times 10^7 \, \nu \mathrm{s^{-1} cm^{-2}} $) from the source. A perfectly efficient detector and negligible neutrino oscillation effects are assumed.}\label{fig.8}
\end{figure}
%%%%%%%%%%%%%%%%%%%%%%%%%%%%%%%%%%%%%%%%%%%%%%%%% %%%%%%%%%%%%%%%%%% 
%
%
%%%%%%%%%%%%%%%%%%%%%%%%%%%%%%%%%%%%%%%%%%%%%%%%%%%%%%%%%%%%%%%%%%%
\begin{figure}[ht]
\begin{center}
\includegraphics[width=1.0\textwidth]{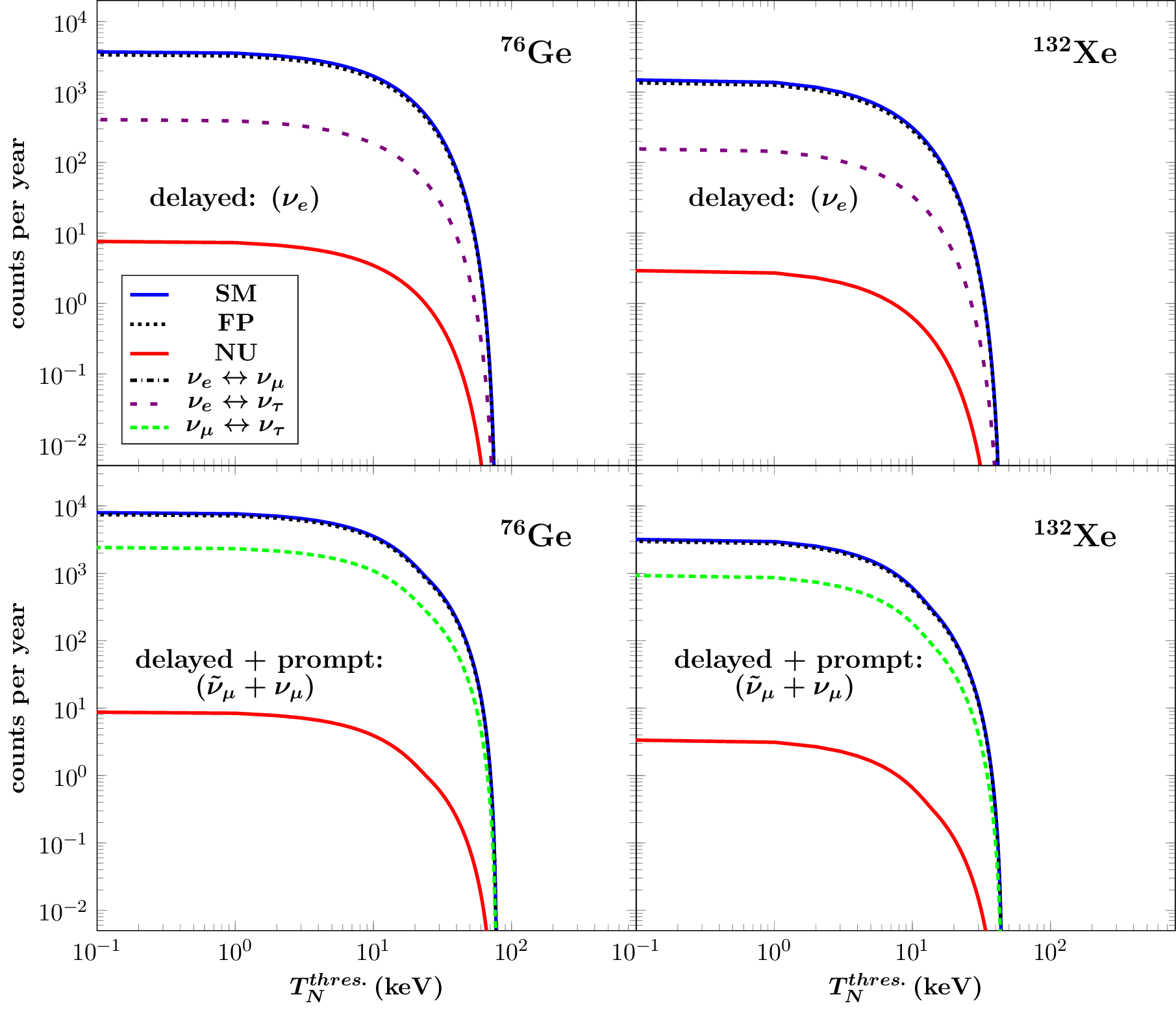}
\end{center}
\caption{Same as Fig. \ref{fig.8}, but for 100 kg of $^{76}$Ge at 20 m ($\Phi_{\nu_{\alpha}}=2.5 \times 10^7\, \nu \mathrm{s^{-1} cm^{-2}} $) and 100 kg of liquid $^{132}$Xe at 40 m ($\Phi_{\nu_{\alpha}}=6.3 \times 10^6\, \nu \mathrm{s^{-1} cm^{-2}} $) from the source.}\label{fig.9}
\end{figure}
%%%%%%%%%%%%%%%%%%%%%%%%%%%%%%%%%%%%%%%%%%%%%%%%%%%%%%%%%%%%%%%%%%% 
%

In Fig. \ref{fig.8} and Fig. \ref{fig.9} the resulting number of exotic events are illustrated and compared with the SM predictions. We note, however, that especially for the case of the flavour changing (FC) channel $\nu_{\mu} \rightarrow \nu_{e}$, by using the extremely high sensitivity of the ongoing $\mu^{-} \rightarrow e^{-}$ conversion experiments (COMET \cite{COMET} and Mu2e \cite{Bernstein-Cooper}), very robust bounds have been set on the vector parameters $\epsilon_{\mu e}^{f V}$ \cite{PLB}. To this end, we conclude that, if the Mu2e and COMET experiments will not detect muon to electron conversion events, then the new $\epsilon_{\mu e}^{f V}$ parameters extracted in \cite{PLB} will lead to undetectable coherent rates at the SNS facility for this channel. 

For our present calculations we use the current bounds \cite{PLB} set by the sensitivity of the PSI experiment \cite{Wintz} and found countable number of events for the near detectors in the case of the corresponding $\nu_{\mu} \rightarrow \nu_{e}$ reaction. The other exotic parameters, i.e. $\epsilon_{\alpha \alpha}^{f V}$ with $\alpha=e, \mu$ and $\epsilon_{e \tau}^{f V}$ have been taken from Ref. \cite{Davidson}. As discussed in Ref. \cite{PLB}, we do not take into account the $\epsilon_{\tau \tau}^{f V}$ contribution, since the corresponding limits are poorly constrained and eventually predict unacceptably high rates.

Before closing, it is worth noting that, the present calculations indicate significant possibility of detecting exotic neutrino-nucleus events through coherent scattering in the aforementioned experiments. Since neutrino-physics enters a precision era \cite{Scholberg}, a difference from the Standard Model predictions leads to an undoubtable evidence of non-standard neutrino-nucleus interactions (NSI). We recall that, in order to experimentally constrain simultaneously all the exotic parameters at high precision, the detector material should consist of maximally different ratio $k=(A+N)/(A+Z)$ \cite{Barranco,Scholberg}. 

Our Future plans include estimation of the incoherent channel which may provide a significant part of the total cross section, especially for energies higher than $E_{\nu} \approx 20-40$ MeV (depending on the nuclear target \cite{Chassioti_2009} and the particle model predicting the exotic process).
%%%%%%%%%%%%%%%%%%%%%%%%%%%%%%%%%%%%%%%%%%%%%%%%%%%%%%%%%%%%%%%%%%%%%%%%%%%%%%%%%%%%%%%%%%%%%%%%%%%%%%%%%%%%%

%%%%%%%%%%%%%%%%%%%%%%%%%%%%%%%%%%%%%%%%%%%%%%%%%%%%%%%%%%%%%%%%%%%%%%%%%%%%%%%%%%%%%%%%%%%%%%%%%%%%%%%%%%%%%
\section{Summary and Conclusions}

Initially, in this paper the evaluation of all required nuclear matrix elements, related to Standard Model and exotic neutral-current $\nu$-nucleus processes is formulated, and realistic nuclear structure calculations of $\nu$-nucleus cross sections for a set of interesting nuclear targets are performed. The first stage involves cross sections calculations for the dominant coherent channel in the range of incoming neutrino-energies $0 \le E_\nu \le 150$ MeV (it  includes $\nu$-energies of stopped pion-muon neutrino decay sources, supernova neutrinos, etc). 

Additionally, new results for the total number of events expected to be observed in one ton of various $\nu$-detector materials are provided and the potentiality of detecting supernova as well as laboratory neutrino-nucleus events is in detail explored. The calculations are concentrated on interesting nuclei, like $^{20}\mathrm{Ne}$ and $^{40}\mathrm{Ar}$, $^{76}$Ge and $^{132}$Xe which are important detector materials for several rare event experiments, like the COHERENT at Oak Ridge National Laboratory, and also experiments searching for dark matter events as the GERDA, SuperCDMS, XENON 100, CLEAN, etc. By comparing our results with those of other methods, we see that the nuclear physics aspects (reflecting the accuracy of the required $\nu$-nucleus cross sections), appreciably affect the coherent $gs \rightarrow gs$ transition rate, a result especially useful for supernova $\nu$-detection probes.   

In the present work, the QRPA method that considers realistic nuclear forces has been adopted in evaluating the nuclear form factors, for both categories of $\nu$-nucleus processes, the conventional and the exotic ones. Also, a comparison with other simpler methods as (i) effective methods and (ii) the method of fractional occupation probabilities, which improves over the simple Shell-Model and gives higher reproducibility of the available experimental data, is presented and discussed. We conclude that among all the adopted methods the agreement is quite good, especially for light and medium nuclear isotopes. However, since coherent neutrino-nucleus scattering can probe the neutron nuclear form factors, methods like the BCS  provide more reliable results.

In view of the operation of extremely intensive neutrino fluxes (at the SNS, PSI, J-PARC, Fermilab, etc.),  the sensitivity to search for new physics will be largely increased, and therefore, through coherent neutrino-nucleus scattering cross section measurements, several open questions (involving non-standard neutrino interactions, neutrino magnetic moment, sterile neutrino searches and others) may be answered. Towards this purpose, we have comprehensively studied the non-standard neutrino-nucleus processes and  provided results for interesting nuclear detectors. Our predictions for the total number  of events indicate that, within the current limits of the respective flavour violating parameters, the COHERENT experiment may come out with promising results on NSI. Moreover, this experiment in conjunction with the designed sensitive muon-to-electron conversion experiments (Mu2e, COMET), may offer significant contribution for understanding the fundamental nature of electroweak interactions in the leptonic sector and for constraining the parameters of beyond the SM Lagrangians.

%%%%%%%%%%%%%%%%%%%%%%%%%%%%%%%%%%%%%%%%%%%%%%%%%%%%%%%%%%%%%%%%%%%

\section{Acknowledgements}
One of us, (DKP) wishes to thank Dr. O.T. Kosmas for technical assistance.  

%% References


\section{References}
\bibliographystyle{elsarticle-num}

%
\vspace{10.0cm}

%%%%%%%%%%%%%%%%%%%%%%%%%%%%%%%%%%%%%%%%%%%%%%%%%%%%%%%%%%%%%%%%%%% 
%
%
%%%%%%%%%%%%%%%%%%%%%%%%%%%%%%%%%%%%%%%%%%%%%%%%%%%%%%%%%%%%%%%%%%%
%
\end{document}